\documentclass[]{aa} 

\usepackage[switch]{lineno}

\usepackage[colorlinks=true, linkcolor=blue, citecolor=blue, urlcolor=blue]{hyperref}
\makeatletter
\renewcommand*\aa@pageof{, page \thepage{} of \pageref*{LastPage}}
\makeatother

\usepackage[varg]{txfonts}
\usepackage{amsmath,amssymb,mathrsfs,graphicx,float,latexsym,url}
\usepackage{epstopdf,ragged2e}
\usepackage{natbib,url}
\usepackage{hyperref}

\usepackage[usenames,dvipsnames]{color}
\definecolor{darkred}{rgb}{0.5,0,0}
\definecolor{darkgreen}{rgb}{0,0.5,0}
\definecolor{darkblue}{rgb}{0,0,0.5}
\definecolor{prussian}{rgb}{0.0, 0.19, 0.33}
\definecolor{richelectricblue}{rgb}{0.03, 0.57, 0.82}
\definecolor{teal}{rgb}{0.0, 0.5, 0.5}
\definecolor{mediumseagreen}{rgb}{0.24, 0.7, 0.44}
\definecolor{lust}{rgb}{0.9, 0.13, 0.13}
\definecolor{ballblue}{rgb}{0.13, 0.67, 0.8}
\definecolor{darkcyan}{rgb}{0.0, 0.55, 0.55}
\definecolor{mountainmeadow}{rgb}{0.19, 0.73, 0.56}
\definecolor{palecarmine}{rgb}{0.69, 0.25, 0.21}
\definecolor{richcarmine}{rgb}{0.84, 0.0, 0.25}
\definecolor{tangelo}{rgb}{0.98, 0.3, 0.0}
\definecolor{venetian}{rgb}{0.784,0.031,0.082}
\definecolor{bdfrance}{rgb}{0.192,0.549,0.906}
\usepackage{mathtools}
\usepackage{amsbsy}
\usepackage{bm}
\usepackage{float}

\newcommand{\fGW}{f_{\rm GW}}
\newcommand{\Mc}{M_{\rm c}}
\newcommand{\Ms}{M_{\rm WD}}

\newcommand{\be}{\begin{equation}}
\newcommand{\ee}{\end{equation}}
\newcommand{\bear}{\begin{eqnarray}}
\newcommand{\eear}{\end{eqnarray}}

\newcommand{\fgw}{f_{\rm GW}}
\newcommand{\fdgw}{\dot{f}_{\rm GW}}

\def\apj{{ApJ}}
\def\aj{{AJ}}
\def\apjs{{The Astrophysical Journal Supplement}}
\def\apjl{{ApJL}}

\def\aap{{A\&A}}

\def\mnras{{MNRAS}}

\def\nat{{Nature}}

\def\prd{{Physical Review D}}

\def\04a{{2004 a}}
\def\04b{{2004 b}}

\begin{document}

   \title{Population synthesis and detection prospects for Galactic long-period transients with LISA}

   \author{Arthur G. Suvorov 
          \inst{1}, Nikolaos Karnesis \inst{2,3,4}, and Valeriya Korol \inst{5}}
  \institute{\inst{1} Theoretical Astrophysics, IAAT, University of T{\"u}bingen, T{\"u}bingen, D-72076, Germany\\
            \email{a.suvorov@uni-tuebingen.de}  \\\inst{2} Department of Physics, Aristotle University of Thessaloniki, Thessaloniki 54124, Greece \\
            \inst{3} Institute for Astronomy, Astrophysics, Space Applications and Remote Sensing,  National Observatory of Athens, 15236 Penteli, Greece 
            \\
            \inst{4} Research Center for Astronomy and Applied Mathematics, Academy of Athens, Soranou Efessiou 4, 11527 Athens, Greece\\\email{nkarnesis@academyofathens.gr} \\
            \inst{5} SRON Space Research Organisation Netherlands, Niels Bohrweg 4, 2333 CA Leiden, the Netherlands\\\email{v.korol@sron.nl}}

   \date{Received ??; ??}

\abstract{The recently-discovered long-period radio transients represent a puzzling new class of astrophysical sources, some of which are thought to be compact binary systems emitting a pulse once per orbit. 
If this interpretation is correct, the orbital period—and therefore the gravitational-wave frequency—is directly encoded in the radio signal, which typically lies in the millihertz band. 
In this work, we explore whether these systems can be detected by the Laser Interferometer Space Antenna (LISA). 
Assuming that this phase-locking between radio pulses and orbital motion applies broadly across the population, we construct synthetic source catalogues informed by both observations and theoretical models of related systems, such as cataclysmic variables. 
We estimate that between $\sim 0.05\%$ and $\sim 6\%$ of long-period transients will be detectable within four years of observation with LISA, depending on astrophysical assumptions. 
For detectable systems, we find injected frequencies are recoverable to one part in $\sim 10^{5}$, amplitudes to within a factor $\sim 2$, and the sky positions to within $\sim$~30 square degrees. 
Our results demonstrate that gravitational-wave observations can provide direct evidence for the binary nature of these sources and, importantly, can guide future radio surveys by predicting pulse periods, sky positions, and orbital properties.
The mismatch between extracted frequency derivatives and that imposed by gravitational-wave decay is also computed to show that the likelihood of electromotive losses driving orbital evolution can be assessed for each detectable candidate.
}

   \titlerunning{Identifying LPTs with LISA}
   \authorrunning{A. G. Suvorov, N. Karnesis, V. Korol}
   \maketitle

\section{Introduction} 
\label{sec:intro}

Long-period transients (LPTs) are a class of compact sources that pulse coherently in the radio band with periods ranging from several minutes to a few hours\footnote{See \url{https://lpt.mwa-image-plane.cloud.edu.au/} and \url{https://vast-survey.org/LPTs/} for living catalogues of sources.}, significantly longer than the millisecond-to-second periods typical of radio pulsars. 
Their nature is a mystery owing to a plethora of observed periodicities, polarisation patterns, and multiband emissions \cite[see][for a review]{rea26}.
At least some LPTs consist of compact binaries involving a white dwarf (WD) primary with the pulse period equalling the orbital period \citep{hw24,rui24}. 
The periodic emission cycle in these systems hints that the mutual magnetosphere within an interaction zone is frozen relative to the orbit such that pulses can consistently reach the observer once per revolution \citep{qu25}.

It has recently been suggested by \citet{yang25} that {many} LPTs represent a stage of binary evolution preceding the cataclysmic variable (CV) phase, in which a WD accretes material from a non-degenerate low-mass companion star. 
In this scenario, the system is in a ``pre-polar'' state, prior to the onset of significant mass transfer, allowing coherent radio emission—potentially via a maser process—to escape without being choked \citep{zhong25}. 
As the system evolves, it may eventually transition into an accreting phase \cite[see also][]{bloot25}.
This scenario is supported by the overlap of observed orbital periods between LPTs and CVs, with the key idea being that strong magnetic fields enable radio emission while preventing disc formation. 
Other scenarios involving magnetars \citep{coop24,lan26}, neutron stars accreting from the interstellar medium \citep{afon23,afon24}, exotic binaries of black holes and compact stars \citep{xiao24}, or even isolated WDs \citep{hw22,loeb22} are not yet definitively ruled out for some sources though. 

{One major reason for so many theoretical proposals is that of a wide spread of LPT periods. 
In fact, many systems show pulsations with periods less than 80~min. 
Such a value lies below the orbital-period limit set by standard evolution theory for (magnetic) CVs where the secondary becomes degenerate, mass loss causes it to expand, and the orbit must widen through a ``period bounce'' \citep{pac81}.
This implies that, in the scenario where pulses are roughly phase-locked to the orbit -- as is the case for the LPTs ILT J1101+5521 \citep{rui24}, GLEAM-X J0704--37 \citep{hw24} and ASKAP J174508.9--505149 \citep{rose26} -- there must be (at least) two subpopulations.
Alternatively, it could be that pulsations in shorter-period LPTs are associated with the intrinsic spin of the primary \citep{bloot25}.}

To constrain the nature of LPT progenitors, we explore complementary observational channels beyond the radio band.
As noted by \cite{suv25}, those LPTs with periods $P \lesssim 1$~hr are actually promising sources for space-based gravitational wave (GW) interferometers \cite[see also][]{zhan26}. 
This is because -- if indeed they involve phase-locked binaries with the pulse period matching the orbit -- these systems would be bright in the band of the Laser Interferometer Space Antenna \cite[LISA;][]{amaro23} and other instruments sensitive at $\sim$~mHz GW frequencies \citep[see, e.g.,][]{TianQin,Taiji}.
Searching for LPTs in the LISA data stream thus provides a path to confirm binary scenarios (involving WDs or otherwise).

In this paper, we construct synthetic populations of LPTs by drawing from distributions motivated through a combination of theoretical expectations and/or properties of the observed population.
{We focus on an approximately phase-locked model, and ask specifically to what degree \emph{that subpopulation} would be detectable with LISA.
To maintain self-consistency in this respect, we consider evolved, degenerate companions for LPTs in tight orbits and adopt a ``pre-polar'' model involving M dwarfs in wider ones following \cite{yang25}. 
Importantly, we do not necessarily claim an evolutionary link between these two subsets but rather impose it at a phenomenological level to avoid assigning physically inconsistent hydrogen-rich companions to ultra-short-period systems.}
Using established pipelines for parameter estimation and source recovery from injections into a hypothetical LISA data stream \citep{korol22,geo23,katz25}, we aim to not only estimate the fraction of LPTs that are visible to LISA but how GW data can be combined with those in the electromagnetic bands to pinpoint their characteristics (e.g., the nature of binary components, orbital-decay mechanism, and distribution on the sky).

We also turn this question around and explore whether a compact binary detected by LISA could be identified as an LPT. One possible approach is to note that some form of energy loss is required to power the observed radio emission.
In the pre-CV scenario, this energy must ultimately be drawn from the orbit, implying a departure from the standard GW-driven evolution \citep{peters64}.
If one can measure the orbital frequency derivative together with the chirp mass well enough, estimates for the mismatch can be made and compared to electromotive braking formulae \citep{lai12}.
We find in fact that LISA may be used to guide radio surveys, as the sky localisation can be narrowed down {to $\mathcal{O}(\text{few})$~square-degrees} for systems with signal-to-noise (SNR) ratios exceeding $\sim 10$.
Such a localisation is within the field of view (FOV) for many of the radio telescopes that have detected LPTs \citep{murph26}.
Indeed, as emphasised by \cite{rea26} and others, it is likely that the \emph{currently observed} population is a small subset of the total cohort. 
Aside from using LISA to unravel their nature, it may therefore be independently important to account for an LPT foreground (the combined signal from low-SNR binaries) that could manifest in mHz gravitational wave surveys, as for CVs \citep{scar23} and other compact populations \citep{mcm26}.

This paper is organised as follows. 
We review pertinent properties of LPTs in Section~\ref{sec:lpts} to set the stage for population synthesis based on either extrapolating from the existing sample or from theoretical expectations in the pre-CV picture. 
The building of catalogues involves constructing distributions for their periods (Sec.~\ref{sec:sample}), source density in the sky (Sec.~\ref{sec:populationnumber}), frequency derivatives (Sec.~\ref{sec:periodderivs}), and component masses (Sec.~\ref{sec:chirpmasses}) .
Equipped with these, we describe the anticipated GW luminosities in Sec.~\ref{sec:gws} and the pipelines used to quantify SNRs (Sec.~\ref{sec:signaltonoise}).
Results are presented in Sec.~\ref{sec:results} with respect to source resolvability (Sec.~\ref{sec:resolve}), the accuracy with which injected parameters are recovered (Sec.~\ref{sec:parameterrecovery}), and multimessenger connections (Sec.~\ref{sec:outlook}). 
Discussion and conclusions are given in Sec.~\ref{sec:conclusions}.

\section{Phase-locked binaries as long-period transients} 
\label{sec:lpts}

\begin{figure}
\centering
 \includegraphics[width=0.48\textwidth]{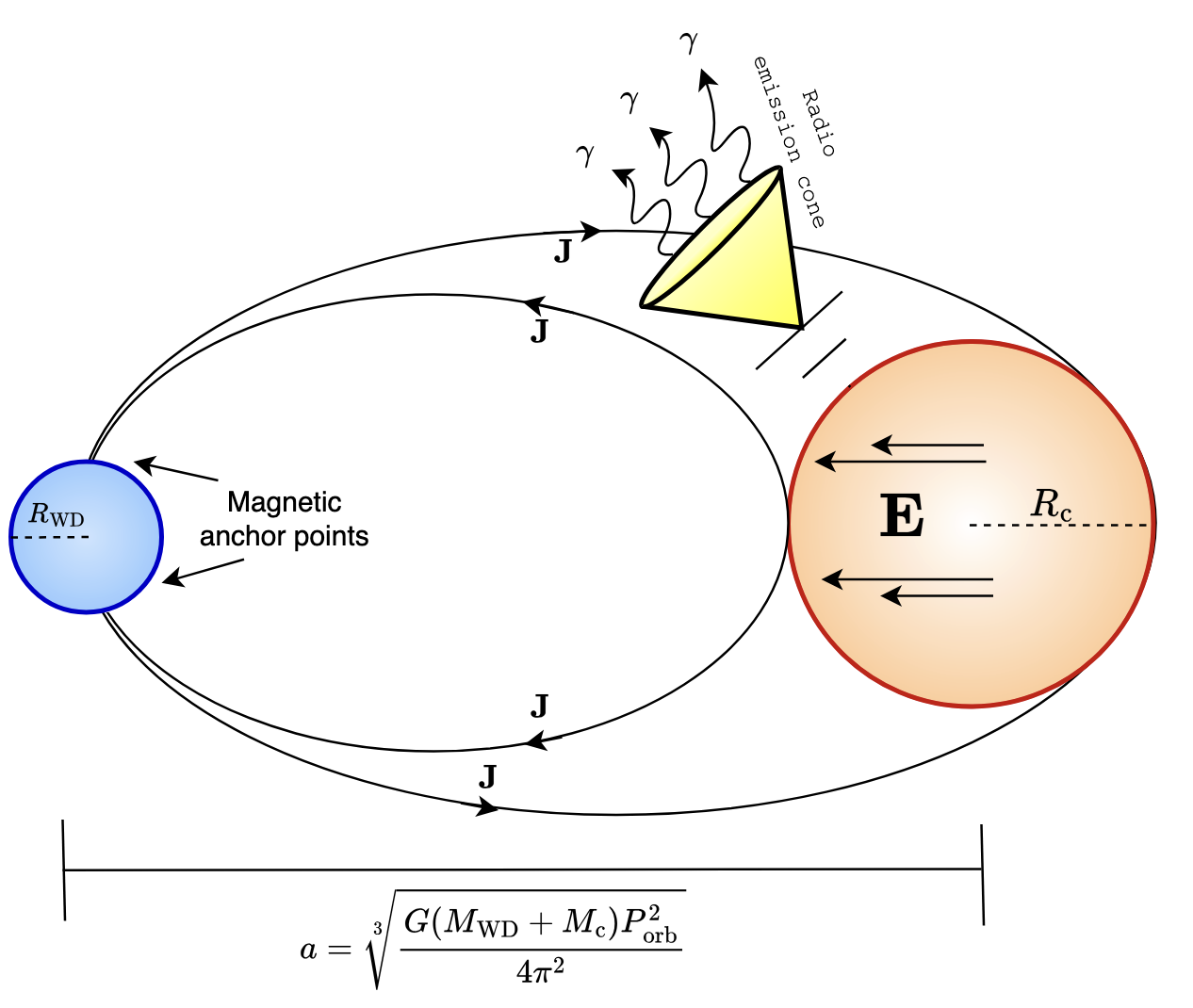}
 \caption{Schematic of the unipolar inductor in a phase-locked, binary LPT consisting of a white dwarf, of radius $R_{\rm WD}$, and a companion, of radius $R_{\rm c}$, in a circular orbit with separation $a$. 
 Plasma currents, $\boldsymbol{J}$, are driven within a flux tube circuit, with cyclotron resonances preferentially occurring in sections of the tube near the companion \cite[see figure 2 in][]{zhong25}.
 An induced electromotive force contributes to orbital decay, perhaps dominantly if the electric field, $\boldsymbol{E}$, is large (Sec.~\ref{sec:periodderivs}).}
  \label{fig:schematic}
\end{figure}

As already noted, there are a number of models that can theoretically explain LPT characteristics \cite[see][for an overview]{rea26}.
This is because many distinct mechanisms, in principle, can instigate transient radio pulsing with periodicities ranging from tens of minutes to hours, either from an isolated object or from a binary.
Their brightness temperatures ($\gtrsim 10^{16}$~K) demand compact sources, but as most LPTs lack multiband characteristics it is difficult to pinpoint their nature. 

In this work, we adopt the picture that LPTs consist of binaries involving (at least one) WD. 
This is motivated by the fact that some sources have been verified as such through optical data and, even in cases without direct confirmation, boast properties that suggest an evolutionary link to CVs \cite[see, e.g.,][]{rid23,rod25}. 
Notably, \cite{yang25} argues that many LPTs are binaries in which strong magnetic fields prevent the formation of an accretion disc or disrupt an existing one.
Strong fields could permit radio pulsing even if the binary resides within the X-ray ``period gap'' associated with CVs ($P_{\rm orb} \lesssim 80$~min), when the donor star no longer fuses in its core, the system detaches, and X-ray emissions shut off \citep{scar23}. 

{It is important to note though that if $P_{\rm pulse} \approx P_{\rm orb}$, LPTs with $P_{\rm pulse} \lesssim 80$~min are unlikely to have main-sequence companions due to period bounce phenomena \citep{pac81}.
A more promising category for these sources involves a primary WD with a (semi-)degenerate companion, more akin to the AM CVn stars\footnote{{We remark in passing that there is some evidence of radio activity in the prototypical AM CVn system HM Cancri for which the orbital period is $\approx$~5.4 min \citep{ram07} and that the system is near-synchronised \citep{dall07}.}}.
We account for the equations of state self-consistently in building catalogues, as described in subsequent sections.}

{Given that compact binaries are expected to be strong sources of GWs for space-based interferometers,} this prompted \cite{suv25} to examine the prospect of searching for LPTs not only with electromagnetic but also with gravitational-wave radiation. 
The results are promising: they found that several members of the existing population, such as ILT/CHIME J1634+44 \citep{bloot25,dong25} and GLEAM-X J162759.5-523504.3 \citep{hw22}, may be GW-visible with $\text{SNR} \gtrsim 10$ within a 4-year observation under reasonable assumptions.
In this work, we extend that study by constructing synthetic catalogues of LPTs, based on either the existing sample or from theoretical considerations. 
At the very least, GW data can be used to identify if LPTs consist of tight binaries and thus rule out models involving isolated stars.

{We rely on two major assumptions in this paper, being that of (i) no active accretion and (ii) phase-locking. 
These are discussed in the next sections.} 

\subsection{No active accretion} \label{sec:roche}

In a pre-polar or binary model generally, a lack of active accretion is crucial to explain radio onset as coherent emissions from a cyclotron maser require that pulses not be choked by circling plasma \citep{qu25,zhong25}. 
The basic mechanism involves the formation of a unipolar inductor (UI), where a flux tube between the closed field-line bundles sourced by the magnetised primary connects the two orbiters, with resonances between electrons and electromagnetic waves instigating radio pulsing \citep{gold69}.
Such models have been studied thoroughly in the context of radio emissions from planets and their satellites, where the observed periodicities are strongly correlated with the orbital one \citep{kav23,zhong25}.
A schematic of the scenario is depicted in Figure~\ref{fig:schematic}.

It is still possible though for Roche-Lobe overflow to occur in these systems provided the induced electric field that works to accelerate particles is sufficiently large relative to that set by the internal plasma screening; as estimated by \cite{yang25}, this caps the accretion rate to\footnote{The radio cycle could, in principle, be linked to state-switching in the system with disk or magnetoviscous instabilities sporadically depleting the inner regions, allowing for transient pulsing, even if the accretion rate sometimes exceeds $\dot{M}_{\rm max}$ \citep{sg26}.}
\begin{equation} \label{eq:mdotcond}
    \dot{M}_{\rm max} \lesssim 10^{-10} \left(\frac{\mu_{\rm WD}}{10^{34} \text{ G cm}^3}\right)^{2} \left(\frac{P_{\rm orb}}{90 \text{ min}}\right)^{-{11}/{3}} M_{\odot} \text{ yr}^{-1},
\end{equation}
for ``canonical'' orbits and masses (e.g., $M_{\rm WD} \approx 0.8 M_{\odot}$; see equation 20 therein), where $\mu_{\rm WD}$ is the dipole moment of the primary.
As the surface temperature of an accreting, mature WD is related to the mass flux through $T_{\rm eff}^4 \propto \langle \dot{M} \rangle$ \citep{towns03}, the above picture is consistent with the lack of \emph{persistent} X-rays from most LPTs.
Small hotspots could form however by the dissipation of currents near the magnetic anchor points at the WD surface and could lead to X-ray emissions, as for ASKAP J144834--685644 \citep{ak25} {and ASKAP J174508.9--505149 \citep{rose26}} (see Fig.~\ref{fig:schematic}).

{Nevertheless, we consider a more conservative scenario in this work where we enforce a lack of Roche-lobe overflow.
This requires the radius of the companion, $R_{\rm c}$, to remain strictly inside its volume-equivalent Roche-lobe radius, $R_L$, as given by the \cite{egg} formula,
\begin{equation} \label{eq:radiuscond}
    R_{\rm c} < R_L = a \times \frac{0.49 q^{2/3}}{0.6 q^{2/3} + \ln\left(1 + q^{1/3}\right)},
\end{equation}
where $q = M_{\rm c}/M_{\rm WD}$ is the mass ratio.
Thus, assuming an equation of state to relate $M_{\rm c}$ to $R_{\rm c}$ and Keplerian motion with 
\begin{equation} \label{eq:kepler}
    a^3 = G( M_{\rm WD} + M_{\rm c}) P_{\rm orb}^{2}/4\pi^2,
\end{equation}
an implicit constraint on $R_{\rm c}$ follows.
The mass-radius relation of the secondary depends on its (in general unknown) composition.
As those with LPTs with $P_{\rm orb} \gtrsim 80$~min and confirmed binarity have M dwarf companions \citep{hw24,rui24,imbrog26,rose26} even if not phase-locked \citep{marsh16,peli24}, it is reasonable to posit this for the longer-period members.
In this case, we adopt an approximate mass-radius relation for M/K dwarfs in the form \citep{kippen94}
\begin{equation} \label{eq:mdmrreln}
    R_{\rm c}/R_{\odot} \approx 0.83 \left(M_{\rm c}/M_{\odot}\right)^{0.84}.
\end{equation}
We incorporate a flag which caps the ceiling value of $M_{\rm c}$ to $0.6 M_{\odot}$ corresponding to the threshold for the M/K spectral transition \cite[$\sim 3800$~K; see, e.g.,][]{raj18}. 
Raising this ceiling would lead to more optimistic predictions without changing the main conclusions of this work (in line with our attempt to be \emph{conservative}).}

{If instead $P_{\rm orb} \lesssim 80~$min, we anticipate degenerate companions \citep{pac81,scar23}. At this stage we emphasise that such a switch is not intended to describe a continuous evolutionary track, but to prevent physically inconsistent hydrogen-rich companions from being assigned to ultra-short-period systems.
We use the analytic, zero-temperature model from \cite{nau72},
\begin{equation} \label{eq:altmreln}
 R_{\rm c}/R_{\odot} \approx  \frac{0.0225}{\mu} \times\frac{1 - (M_{\rm c}/\tilde{M})^{4/3}}{(M_{\rm c}/\tilde{M})^{1/3}},
\end{equation}
where $\tilde{M} = 5.816 M_{\odot}/\mu^2$ is a constant depending on the effective chemical potential, $\mu$. 
Assuming helium companions, we have $\mu = 2$.
When building catalogues, a flag is introduced to identify whether a drawn orbital period lies above $80$~min in which case relation \eqref{eq:mdmrreln} is used, else expression \eqref{eq:altmreln} is used, and values of $R_{\rm c}$ are considered such that condition \eqref{eq:radiuscond} is enforced as an upper limit.
Importantly, if we were to instead focus primarily on double degenerates (i.e., use equation \ref{eq:altmreln} for any $P_{\rm orb}$), detection prospects improve \citep{zhan26} and thus our model can be considered \emph{conservative}.}

\subsection{Phase locking} \label{sec:phaselocking}

To move forward, we operate under the assumption that the pulsation period matches that of the orbit.
This is the case for the confirmed-binaries ILT J1101+5521, GLEAM-X J0704--37, {and ASKAP J174508.9--505149} though is notably \emph{not} for other objects that could retrospectively be considered LPTs, such as  Ar Scorpii \citep{marsh16} or its siblings J191213.72--441045.1 \citep{peli24} and SDSS J230641.47+244055.8 \citep{castro25}. 
{The reason for the discrepancy is likely due to the degree of asynchronicity between the spin period of the primary and the orbit set by the efficacy of electromotive torques.
This is consistent with the fact that other ``WD pulsars'' are between 10 and 100 times fainter in terms of radio luminosity compared to LPTs that are confirmed to be phase-locked \citep{rod25}: for strong fields ($\mu_{\rm WD} \gtrsim 10^{34} \text{ G cm}^3$), even a small degree of asynchronicity could be sufficient}.
While estimates provided in this paper may thus not apply to the entire LPT population, assuming $P_{\rm pulse} \approx P_{\rm orb}$ provides a reasonable working hypothesis. 

{In general, the rotation of the primary tracks the shrinking orbit via the flux tube connecting it to the companion (see Fig.~\ref{fig:schematic}). 
To remain tightly synchronised, the fractional slip, $\delta = |P_{\rm orb}-P_{\rm spin}|/P_{\rm orb}$, should be sufficiently small such that $P_{\rm orb} \approx P_{\rm spin}$ so that pulsations are unambiguously tied to the orbital period \cite[see, e.g.,][]{dall07,bloot25}.
In a steady-state inspiral, the UI torque must continuously counteract the inertial lag of the primary to force $\dot{\Omega}_{\rm spin} \approx \dot{\Omega}_{\rm orb}$. 
In general, the required acceleration torque is
\begin{equation} \label{eq:treq}
T_{\rm req} = I_0 \dot{\Omega}_{\rm orb} = I_0 \pi \dot{f}_{\rm GW},
\end{equation}
where 
\begin{equation} \label{eq:gwfreq}
    \fGW = 2/P_{\rm orb}
\end{equation}
is the GW frequency and $I_{0}$ is the moment of inertia of the primary.
Because the WD is spinning slowly compared to its breakup speed in such a scenario ($P_{\rm spin,max} \sim 10~$s), the rotational deformation (oblateness) is negligible, and we can again adopt expression \eqref{eq:altmreln} for the mass-radius relation with $\mu \approx 2$ and $I_{0} \approx 0.2 M_{\rm WD} R_{\rm WD}^2$ \citep{st83}. }

{There are a number of models for the UI torque available in the literature, primarily owing to uncertainties about the effective resistance in the magnetospheric interaction zone and the field strength of the companion \cite[see, e.g.,][]{skk24}.
Assuming the dissipation occurs via Alfv{\'e}nic flux tubes tracking the cross-section of an unmagnetised secondary, the relevant torque can be modelled as \citep{gold69,lai12}
\begin{equation} \label{eq:uit}
T_{\rm UI} \approx \frac{\mu_{\rm WD}^2 R_{\rm c}^2}{2a^5} \delta.
\end{equation}
Note that $T_{\rm UI}$ vanishes as $\delta \to 0$ because the flux tube would be perfectly co-rotating with the orbit and therefore no work is done.
Again assuming Keplerian motion \eqref{eq:kepler}, we have
\begin{equation} \label{eq:tui}
T_{\rm UI} = \frac{\pi^{10/3}\mu_{\rm WD}^2 R_{\rm c}^2   f_{\rm GW}^{10/3}}{2 G^{5/3} \left(M_{\rm WD} + M_{\rm c}\right)^{5/3}} \delta.
\end{equation}
Equating the required inertial torque \eqref{eq:treq} to the magnetic torque \eqref{eq:tui} and isolating the radius yields
\begin{equation} \label{eq:mastertracking}
    R_{\rm c}^2 = \frac{2 I_0 G^{5/3} (M_{\rm WD} + M_{\rm c})^{5/3} \dot{f}_{\rm GW}}{\pi^{7/3} \fGW^{10/3} \mu_{\rm WD}^2 \delta} .
\end{equation}
If we permit a maximum slippage,
\begin{equation}
    \delta \leq \delta_{\rm max},
\end{equation}
equation \eqref{eq:mastertracking} places either a lower or upper limit on $M_{\rm c}$ if the companion is an M dwarf or degenerate, respectively, as either the mass scales with radius or inversely.}

{In general therefore, enforcing a lack of active accretion and permitting a small maximum slippage ($\delta_{\rm max} = 0.01$) to maintain phase-locking bounds the range of plausible companion masses. 
Sampling from the allowed range allows us to construct chirp masses, which will form the basis for amplitude synthesis for GWs (see Sec.~\ref{sec:gws}).
Note that such a $\delta_{\rm max}$ lies comfortably within the maximum offset permitted by narrowband GW searches \citep{abb08}.
Henceforth, we drop subscripts on the symbol $P$ where no ambiguity can arise since the orbital and pulse periods are equal up to factors of $\approx 1\pm\delta$ in the scenario put forward here.}

\section{Population synthesis} \label{sec:popsynth}

\subsection{Observational sample and period distribution} 
\label{sec:sample}

\begin{table}
\centering
\caption{Inferred periods (in descending order) and period derivatives for the LPTs forming the base sample for Population I used in this work (some names have been shortened; we refer to the cited references and the catalogue in Footnote 1 for the full astronomical denominations).}
\begin{tabular}{lcc}
 \hline
 \hline
Source name & $P$ (s) & $|\dot{P}|$ (s/s; $1 \sigma$ limit) \\
\hline
CHIME J0630+25$^{\text{a}}$ & 421.36 & $7.8(1.4) \times 10^{-13}$  \\
ILT/CHIME J1634+44$^{\text{b}}$ & 841.25  & $9.03(11) \times 10^{-12}$  \\
GLEAM-X J1627$^{\text{c}}$ & 1091.17 & $<1.2 \times 10^{-9}$ \\
GPM J1839--10$^{\text{d}}$ & 1318.20 & $3.6(30) \times 10^{-13}$ \\
ASKAP J1832$^{\text{e}}$ & 2656.25 & $<9.8 \times 10^{-10}$ \\
ASKAP J1935$^{\text{f}}$ & 3225.31 & $<2.7 \times 10^{-10}$ \\
ASKAP J1755$^{\text{g}}$ & 4186.33 & $< 1.0 \times 10^{-10}$ \\
GCRT J1745--3009$^{\text{h}}$ & 4627.8 & -- \\
ASKAP J1448$^{\text{i}}$ & 5631.07 & $< 2.2 \times 10^{-8}$ \\
ILT J1101+5521$^{\text{j}}$ & 7531.79 & $<1.7 \times 10^{-11}$ \\
GLEAM-X J0704--37$^{\text{k}}$ & 10496.56 & $<1.3 \times 10^{-11}$\\
Ar Scorpii$^{\text{l}}$ & 12833.4 & $<3.3 \times 10^{-10}$ \\
J1912–4410$^{\text{m}}$ & 14522.4 & --\\
\hline
\hline
\end{tabular}
\\
\justify \noindent \footnotesize{{Notes.} $\vphantom{\text{a}}^{\text{a}}$Adopting the no-glitch fit; \cite{dong24} $\vphantom{\text{b}}^{\text{b}}$Two orbital periods are plausible owing to interpulse and spin-orbit phenomena. We consider the shorter here for concreteness; \cite{bloot25,dong25}  $\vphantom{\text{c}}^{\text{c}}$\cite{hw22} $\vphantom{\text{d}}^{\text{d}}$\cite{hw23} $\vphantom{\text{e}}^{\text{e}}$\cite{wang24} $\vphantom{\text{f}}^{\text{f}}$\cite{caleb24} $\vphantom{\text{g}}^{\text{g}}$\cite{sween25} $\vphantom{\text{h}}^{\text{h}}$\cite{hyman09} $\vphantom{\text{i}}^{\text{i}}$\cite{ak25} $\vphantom{\text{j}}^{\text{j}}$3$\sigma$ limit on $\dot{P}$; \cite{rui24} $\vphantom{\text{k}}^{\text{k}}$\cite{hw24}. $\vphantom{\text{l}}^{\text{l}}$Using the orbital, rather than beat, period; \cite{marsh16,stiller18}. $\vphantom{\text{m}}^{\text{m}}$As for Ar Scorpii; \cite{peli23,peli24}.}
\label{tab:lptdata}
\end{table}

Our first task is to build a synthetic sample of LPTs to eventually investigate what fraction may be detectable in GWs. 
We begin by briefly reviewing pertinent aspects of the known LPT population to make informed choices for the relevant statistical priors.

Table~\ref{tab:lptdata} shows relevant period and period-derivative data for the LPT population which are viably binaries\footnote{{As this work was being completed two new LPTs, ASKAP J142431.2--612611 and ASKAP J174508.9-505149, were discovered with $P \approx 35.8$~min \citep{prit26} and $P \approx 82.1$~min \citep{rose26}, respectively. Inclusion of these source within the empirical sample would lead to \emph{more optimistic} estimates owing to their relatively short periods.}}.
In particular, as stressed earlier, it could be the case that LPTs consist of multiple subclasses. 
For instance, the detection of interpulses with a phase offset of $\approx 0.5$ in ASKAP J183950.5--075635.0 points strongly towards emissions from both magnetic poles of a neutron star \citep{lee25}. 
It is for this reason we have excluded this and a small number of other sources, as discussed in \cite{suv25}: our criteria are the same as that adopted there. 

\begin{figure}
\centering
 \includegraphics[width=0.487\textwidth]{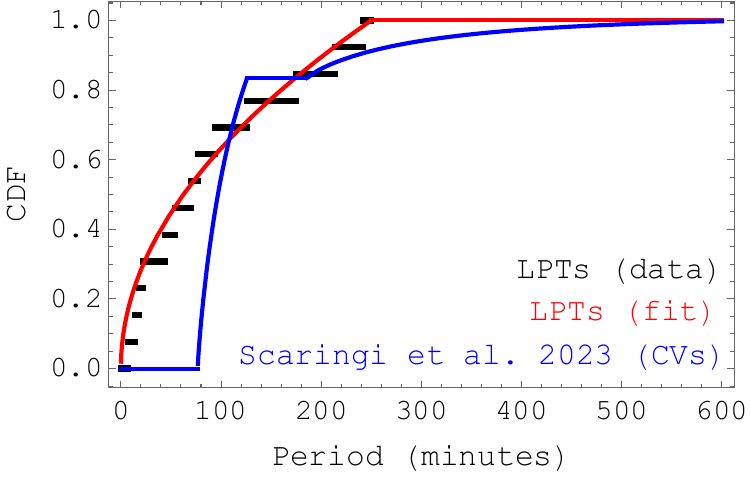}
 \caption{Empirical CDFs for the observational sample of LPTs considered in Tab.~\ref{tab:lptdata} (black), together with a smooth fitting (red). Overlaid in blue is the fit described by \protect\cite{scar23} for CVs; we refer the reader to that work for details (see equation 1 therein).}
  \label{fig:cdfs}
\end{figure}

Because reliable folding of the pulse signal requires many pulses, and LPTs emit sporadically, most of the sources listed in Tab.~\ref{tab:lptdata} have only upper limits on $\dot{P}$.
In any case, the \emph{periods} of the objects in our sample \emph{are} well measured, and we can thus construct an empirical cumulative distribution function (CDF). 
The results are shown in Figure~\ref{fig:cdfs}, together with that associated with CVs as considered by \cite{scar23}. 
It is clear on a visual level that the shape similarity but dearth of CV periods less than $\sim 80$~min gives credibility to the notion that LPTs share a common pathway. 
In particular, if one were to take the CDF considered by \cite{scar23} and ``smear'' it towards lower periods, a significant overlap between the two curves could be expected.

{The CDF for period data in Tab.~\ref{tab:lptdata} is well-fit by a power-law function of the form
\begin{equation} \label{eq:periodcdf}
    \tilde{F}(P) = \frac{P^\alpha - P_L^{\alpha}}{P_{U}^\alpha - P_{L}^{\alpha}}, \quad P_{L} \leq P \leq P_{U},
\end{equation}
with $\alpha \approx 1/2$, where $P_L$ and $P_U$ set the lower and upper-limit cutoffs for the sample.}
We fix these as $P_{L} = 7$~min and $P_U = 4$~hours for concreteness, corresponding roughly to the edges of the known sources. We note that extending $P_U$ would lead to a modest tail of undetectable (in GWs) sources, while lowering $P_L$ would lead to more optimistic estimates; see Sec.~\ref{sec:chirpmasses}.
In order to avoid an unphysical pileup of sources {at the boundaries}, we introduce a Gaussian tapering of the form
\begin{equation} \label{eq:periodcdfA}
    F(P) = \exp\left[\frac{s}{\left(P_U - P_L\right)^2}\right] \tilde{F}(P) \exp\left[{-\frac{s}{\left(P-P_L\right)^2}} \right],
\end{equation}
for some constant $s$ setting the steepness of the tapering (set to 100 in units of $\text{min}^2$). {The prefactor ensures the correct normalisation for the CDF [i.e., $\lim_{P \to P_L} F(P) = 0$ and $F(P_U) = 1$] and a strictly-positive derivative for $\alpha,s >0$.}
The functional form \eqref{eq:periodcdfA} preserves the main features of the power-law \eqref{eq:periodcdf} while avoiding unphysical clusterings.

Selection effects may be contaminating the currently-available sample (i.e., perhaps it is easier to find shorter-period sources).
Moreover, the shorter-period sources may be of a different character {(e.g., with a large slippage $\delta$)}.
For these reasons, we consider a second possibility for the period distribution, where instead it is stipulated that some combination of GW emission and a unipolar inductor are responsible for orbital decay. 
Both of these mechanisms also lead to power-law distributions for the periods.
Depending on which is dominant, a value of $\alpha$, appearing in equation \eqref{eq:periodcdfA} describing the CDF, is effectively selected \cite[see, e.g.,][]{korol22}. 
We focus on the case of UI-driven decay, for which $\alpha \approx 10/3$, for two main reasons: (i) significant electromotive losses may be expected to fuel radio activity \citep{qu25,zhong25}, and (ii) it provides a more conservative population estimate with respect to GW detectability (i.e., larger $\alpha$ means shallower decay and thus a higher density of dim sources).

\begin{figure}
\centering
 \includegraphics[width=0.487\textwidth]{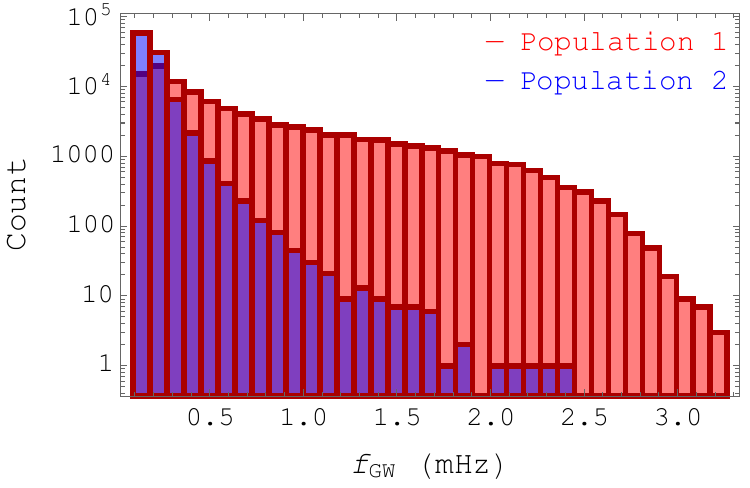}
 \caption{Synthetic sample of $N=10^{5}$ LPT GW frequencies, drawn using the CDF \eqref{eq:periodcdfA} with either $\alpha = 1/2$ (Pop I; red) or $\alpha = 10/3$ (Pop II; blue).}
  \label{fig:fGWlist}
\end{figure}

We thus construct two synthetic populations, referred to throughout as:
\begin{itemize}
    \item[(i)]{{Population I}: from the empirical distribution of the known sources ($\alpha \approx 1/2$; see Fig.~\ref{fig:cdfs}); and} 
    \item[(ii)]{{Population II}: from a distribution corresponding to orbital breaking dominated by electromotive losses ($\alpha \approx 10/3)$.}
\end{itemize}
These two populations represent, in some appropriate sense, the extrema for most optimistic and most conservative scenarios, respectively. {In Appendix~\ref{sec:brokenpower} we demonstrate that using a broken power-law distribution, arguably more physical to separate the two subpopulations at the bounce boundary ($P \approx 80$~min), leads to more optimistic predictions with respect to GWs relative to a UI-dominated track (Population II). 
As such, in an effort to present phenomenological extrema as regards LISA detectability, we consider a single power-law in what follows. 
}

Another necessary ingredient concerns not only the period but the orbital dynamics. To this degree, we assume sources have zero eccentricity; this is motivated by the fact that the eccentricity is expected to be erased long before the systems reach ``short'' periods thanks to GW-radiation reaction \citep{peters64}. This implies that sources radiate only at twice the orbital frequency.
As relevant for LISA, the orbital period distributions considered above are translated into $\fGW$ values from relation \eqref{eq:gwfreq}. 
(Some discussion on how eccentricity adjusts detection prospects via Fourier harmonics is provided in Sec.~\ref{sec:conclusions}). 

Considering a representative total source count of $N= 10^5$ (see Sec.~\ref{sec:populationnumber}), histograms of sampled $\fGW$ values from Pops I and II are shown in Fig.~\ref{fig:fGWlist}. 
We see that the majority of sources appear at large periods and low orbital frequencies; for example, $> 90\%$ of members from Pop II have $f_{\rm GW} < 0.5$~mHz. 
In general, the number of high-frequency sources scales inversely with $\alpha$: over $10\%$ of sources for the $\alpha = 1/2$ case (Pop I) have $f_{\rm GW} > 1$~mHz while only a handful exist in this band for $\alpha = 10/3$ (Pop II). 
This is expected as shallower orbital decay (larger $\alpha$) implies, given some roughly uniform age distribution of sources, longer periods in general. 
Where quantitative estimates are required, these samples are used for the remainder of this paper. 
Some caution is warranted therefore in that differences concerning detection prospects may apply had we gotten ``(un)lucky'' in this particular drawing (given the small number of tail-end sources around $f_{\rm GW} \gtrsim 1$~mHz). 
Nevertheless, given the swathe of uncertainty surrounding LPTs generally and the extremal nature of the two constructed populations, such effects do not significantly affect our conclusions: the true distribution likely resides somewhere between the two.

\subsection{Population density} \label{sec:populationnumber}

As mentioned above, it is important to consider probable numbers for the total number ($N$) of LPTs within the Galaxy. 
Given the reoccurring theme of small samples, together with uncertainties pertaining to activity cycles and selection effects in radio surveys, $N$ is difficult to estimate without extrapolation or evolutionary modelling. 
For instance, the LPT GLEAM-X J1627 was only active for a few months within a multi-year window \citep{hw22} while another, with comparable period, GPM J1839 has been rather persistent over the last three decades \cite[duty cycle $d_{\rm radio} \sim 0.5$;][]{hw23}. 
For concreteness, we follow \cite{yang25} who estimates a number density of $10^{-8} \lesssim n_{\rm LPT}/\text{pc}^{-3} \lesssim 10^{-6}$ (Eq.~108 therein); for a volume of $\sim 10^{12} \text{\,pc}^{3}$, we take the midpoint and fix $N  = 10^{5}$.
{Such a value is also consistent with the population study of \cite{rod26} who estimate $n_{\rm LPT} \sim 10^{-7} \text{ pc}^{-3}$ if radio surveys are $\approx 10\%$ complete.} 

{The adopted value $N=10^5$ should be understood as a fiducial normalisation for the total phase-locked LPT-like population, rather than as a direct prediction from a single binary-evolution channel. 
Systems above and below the canonical CV period minimum are treated as physically distinct: the former are modelled as detached magnetic WD + low-mass-star systems, while the latter require degenerate or semi-degenerate companions. 
Their true Galactic abundances should therefore be bounded by the space densities and radio-active duty cycles of their respective parent populations, such as pre-CV/magnetic-CV-like systems and AM-CVn-like or double-degenerate systems. 
In general, the total number of AM CVn-type systems in the Galaxy is expected to be $\gtrsim 10^5$ \citep{nele04} while for strictly-detached systems the number is vastly higher \citep{nele01}.
Our synthetic populations are thus consistent with the overall number of such sources; for Population I, roughly half of the synthetic sources are such that $P < 80$~min while only $\approx 2\%$ are for Population II.
Nevertheless, we emphasise again that our populations are intended to be phenomenological representations and we do not stipulate an evolutionary link between short and long-period systems; our results are best interpreted as detection efficiencies for a fiducial population and absolute detection numbers scale linearly with the assumed channel normalisations.}

\begin{figure}
\centering
 \includegraphics[width=0.487\textwidth]{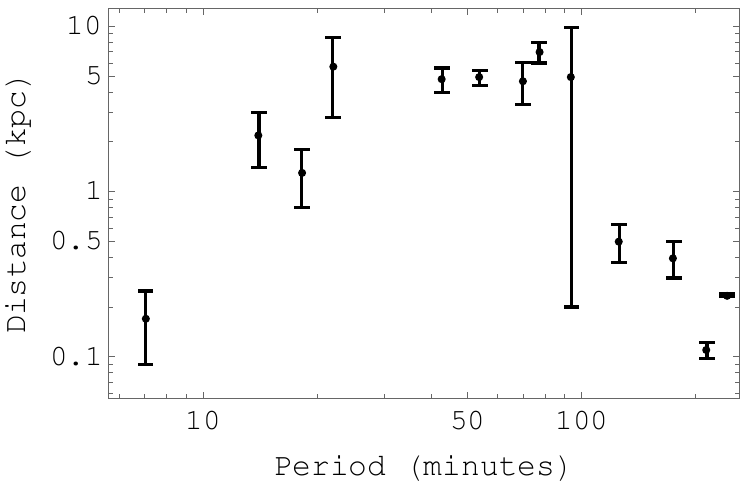}
 \caption{Distance versus pulse period for a sample of the known LPTs, namely those in Tab.~\ref{tab:lptdata}; see Footnote 1 and references therein for details regarding distance uncertainties determined from dispersion measures.}
  \label{fig:distperiod}
\end{figure}

In comparing to the known sources, we note there is no obvious relationship between the distance, $D$, and pulse period. 
This is demonstrated in Figure~\ref{fig:distperiod}, depicting the (dispersion-measure) inferred distances of sources within Tab.~\ref{tab:lptdata} quoted from the respective references.
We see a wide scatter of $P$ and $D$ values (ranging from $\sim 100$~pc to $\sim 10$~kpc). 
Moreover, the Galactic coordinates of the known sources tend to cluster around the centre, though otherwise there is similarly a wide scatter (see Footnote 1). 
While there is no \emph{a priori} reason one may expect any particular relationship between these factors, this motivates the use of uncorrelated distributions for the Galactic positions and $f_{\rm GW}$.

For both populations, we follow \cite{korol22} and others by generating random Galactic coordinates (longitude, latitude) and distances for $N$ objects distributed in a disc, and then transform them to ecliptic coordinates ($\lambda, \ell$).
The radial profile is generated with an exponential-like law (with an exponent of 20/31 and a constant of proportionality of 13 chosen empirically), within the disc distribution using an exponential length-scale of $H_R = 2.5$~kpc, vertical scale-height of $h_{z} = 0.3$~kpc, assuming a distance of 8.1~kpc from Galactic centre to the Sun.
Values of the orbital inclination $\iota$, polarization angle $\psi$, and initial orbital phase $\phi_{0}$ are similarly generated following \cite{korol22}. 
The latter three parameters are inessential for the details considered in this paper; in the context of the LISA literature though, they are often used in the description of quasi-monochromatic, circular binaries.

\subsection{Frequency derivatives and orbital decay} \label{sec:periodderivs}

As is clear from Tab.~\ref{tab:lptdata}, it is difficult to empirically estimate {frequency derivatives for} the known sources as typically only upper-limits for $\dot{P}$ are reported.
This is unsurprising: estimating $\dot{P}$ via folding or otherwise requires many pulses, which are challenging to accrue given the intermittency and long periods of the known sources.  
For power losses orchestrated by GW radiation-reaction we expect, in the limit of a circular binary \citep{peters64},
\begin{equation} \label{eq:pdotgw}
\dot{E}_{\rm GW} = -\frac{256  }{5} \frac{2^{1/3} \pi^{10/3} G^{7/3}}{c^5} \mathcal{M}^{10/3} P^{-10/3},
\end{equation}
which can be translated into a value for $\fdgw$ given values of $P$ and the chirp mass,
\begin{equation} \label{eq:chirp}
\mathcal{M} = \frac {\left( \Ms \Mc \right)^{3/5}} { \left(\Ms + \Mc \right)^{1/5}} ~.
\end{equation}

From the chain rule we can write $dE/dt = dE/dP \cdot dP/dt$ and thus, generically, the orbital evolution can be described through
\begin{equation} \label{eq:orbdecayeqn}
    \frac{\dot{P}}{P} = \frac{3}{2} \frac{( \dot{E}_{\rm GW} + \dot{E}_{\rm UI} )}{E_{\rm grav}},
\end{equation}
with $2E_{\rm grav} = G^{2/3} \mathcal{M}^{5/3} (\pi  f_{\rm GW})^{2/3}$. Noting the scalings $\dot{E}_{\rm GW} \propto P^{-10/3}$ and $\dot{E}_{\rm UI} \sim \Omega_{\rm orb} T_{\rm UI} \propto P^{-13/3}$ from equation \eqref{eq:tui}, we see that electromotive losses lead to a shallower decay of the orbit.

{Rather than appealing to a specific torque formula,} $\dot{E}_{\rm UI}$ can be estimated more directly from the observed radio luminosities of LPTs. 
To do this, consider a radiative efficiency $\epsilon \ll 1$ and write $L_{\rm rad} = \epsilon \dot{E}_{\rm UI}$ \citep{zhong25}.
Both ILT J1101+5521 and GLEAM‑X J0704--37, for instance, show peak radio luminosities of $L_{\rm rad} \sim 10^{28} \text{ erg s}^{-1}$ \citep{rod25} and thus we expect the ratios
\begin{equation} \label{eq:fdotuiILT}
\left(\frac{\dot{E}_{\rm GW}}{\dot{E}_{\rm UI}}\right)_{\rm J1101} \approx 0.68 \times \left(\frac{\epsilon}{2 \times10^{-3}}\right) \left( \frac{\mathcal{M}}{0.29 M_{\odot}} \right)^{10/3},
\end{equation}
and
\begin{equation} \label{eq:fdotuiGX}
\left(\frac{\dot{E}_{\rm GW}}{\dot{E}_{\rm UI}}\right)_{\rm J0704} \approx 0.84 \times \left(\frac{\epsilon}{2 \times 10^{-3}}\right) \left( \frac{\mathcal{M}}{0.43 M_{\odot}} \right)^{10/3},
\end{equation}
respectively. 
Values of $\epsilon \sim 10^{-3}$ are consistent with that coming from kinetic plasma simulations of the cyclotron maser \cite[][see their figure 4]{zhong25}.
If $\dot{f}_{\rm GW}$ can be tracked well enough, the origin of energy losses could thus be identified.

While we could \emph{fix} $\dot{f}_{\rm GW}$ values {directly from $P$ and $\mathcal{M}$ values} by assuming a decay mechanism as above, we opt for a more agnostic approach as $\epsilon$ could vary significantly between systems \citep{lai12} as could the magnitude of other torques \cite[e.g., related to tides, accretion, or propellers; see][]{gs21}.
{Nevertheless, GW losses are likely to be non-negligible, and thus we posit a distribution of frequencies derivatives following}
\begin{equation} \label{eq:fdotdist}
    \fdgw \sim -U(1/b,b) \times \frac{3}{2} f_{\rm GW} \frac{\dot{E}_{\rm GW}}{E_{\rm grav}},
\end{equation}
chosen to be consistent with the scatter in Tab.~\ref{tab:lptdata}, recalling that $\fdgw = -2 \dot{P}/P^2$ from equation \eqref{eq:gwfreq}. 
{In essence, we allow for a scatter around the expected GW-loss value, skewed slightly higher than unity via the constant $b$ to account for electromotive losses. 
We fix $b=5$ for concreteness, though the exact choice is largely unimportant if not too large. Indeed,}
values drawn from \eqref{eq:fdotdist} are consistent with the electromotive loss formulae given above for $\epsilon \lesssim 10^{-2}$.
In any case, as long as $\fdgw$ is sufficiently small -- so as to preserve monochromaticity -- its exact value does not impact significantly on the detectability of any given system.
{In order to generate $\dot{f}_{\rm GW}$ values, however, we require a consistent loop involving the component masses.}

\subsection{Component masses} \label{sec:chirpmasses}

We anticipate that reconnections within the mutual magnetosphere set up by the binary are responsible for radio pulsations from LPTs and thus strong magnetic fields are required \citep{qu25,zhong25}. 
This is consistent with the detection of cyclotron absorption features in the LPT GPM J1839--10 \citep{men26}.
It is thus appropriate to draw primary masses from the \emph{magnetic} subset of WDs.
For the primaries, we consider normal distributions, viz.
\begin{equation} \label{eq:massdistribution}
    M_{\rm WD} \sim N(\mu_{\rm WD}, \sigma_{\rm WD}),
\end{equation}
with $\mu_{\rm WD} = 0.77 M_{\odot}$ and $\sigma_{\rm WD} = 0.1 M_{\odot}$. 
These mean and standard deviations are chosen following the analysis of NuSTAR Legacy Survey data for magnetic WDs by \cite{shaw20}. 

{To estimate companion masses, we make use of the arguments outlined in Sec.~\ref{sec:lpts}. In particular, we draw $N$ values of $\delta$ uniformly from the range $(0,0.01)$ to set the distribution of slippages. From the prepopulated values of $M_{\rm WD}$ and $P_{\rm orb}$ the Roche-lobe condition \eqref{eq:radiuscond} can be checked. 
Combined with $\dot{f}_{\rm GW}$ values, expression \eqref{eq:mastertracking} can then also be evaluated. 
Values for $M_{\rm c}$ are then chosen uniformly from the allowed range of respective limits set by the two conditions, as described in Sec.~\ref{sec:lpts}.
In practice, this involves solving a coupled minimisation problem since both $\dot{f}_{\rm GW}$ and $M_{\rm c}$ feature in the slippage condition \eqref{eq:mastertracking}.
The workflow for either population is as follows. 
(1) Build a list of $P$ values and separate according to whether $P >80$~min or $P< 80$~min. 
(2) Build an array of radii according to equations \eqref{eq:mdmrreln} or \eqref{eq:altmreln} for masses in the range $0.01 < M_{\rm c}/M_{\odot} < 0.6$ for each catagory. 
(3) Identify, via a brute-force grid search, the permitted extrema consistent with the no accretion \eqref{eq:radiuscond} and phase-locking \eqref{eq:mastertracking} conditions such that the sampled $\dot{f}_{\rm GW}$ value is recovered.
(4) Uniformly sample between the extrema, with the selected value defining $M_{\rm c}$ and hence $\mathcal{M}$.
}

\begin{figure}
\centering
 \includegraphics[width=0.487\textwidth]{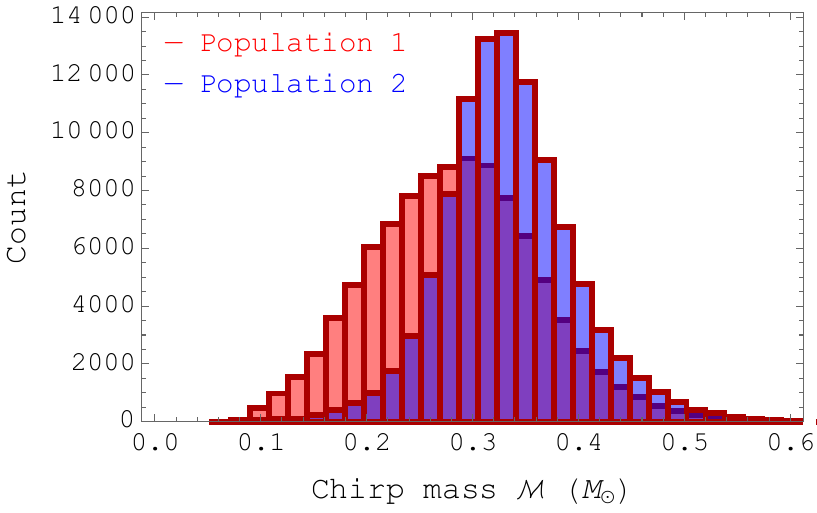}
 \caption{Similar to Fig.~\ref{fig:fGWlist} but showing chirp masses.}
  \label{fig:chirplist}
\end{figure}

Two synthetic histograms for $N=10^{5}$ {chirp masses} drawn from Populations 1 and 2 are shown in Fig.~\ref{fig:chirplist}. 
{Because the Population II objects have systematically lower GW frequencies (Fig.~\ref{fig:fGWlist}), the chirp mass is skewed slightly higher since conditions \eqref{eq:radiuscond} and \eqref{eq:mastertracking} are less restrictive.
The distribution is somewhat flatter for Population I, simply because the period distribution is flatter (see Fig.~\ref{fig:distperiod}) and thus so too is the orbital separation, $a$.
Note that for the confirmed-binary LPTs we have chirp mass estimates of {$\mathcal{M} \approx 0.23 M_{\odot}$ (ASKAP J174508.9--505149)}, $\mathcal{M} \approx 0.29 M_{\odot}$ (ILT J1101+5521), $\mathcal{M} \approx 0.32 M_{\odot}$ (J1912--4410), $\mathcal{M} \approx 0.43 M_{\odot}$ (GLEAM-X J0704--37), and $\mathcal{M} \sim 0.5 M_{\odot}$ (Ar Scorpii); see, for instance, \cite{suv25}. }

{The above estimates are consistent with the mean values shown in Fig.~\ref{fig:chirplist}, which read $\mu_{\mathcal{M}}(\rm Pop \,\, I) \approx 0.29 M_{\odot}$ and $\mu_{\mathcal{M}}(\rm Pop \,\, II) \approx 0.33 M_{\odot}$, with distributions being roughly Gaussian.
These values are slightly higher than -- but again roughly consistent with -- the mean chirp masses inferred from the data compiled in table 2 from \cite{sol10} for AM CVn systems, where $\mu_{\mathcal{M},AM CVn} \approx 0.2 M_{\odot}$ with a standard deviation of $\approx 0.11 M_{\odot}$.
If we were to exclude cases with $P > 80$~min from our samples the chirp-mass distribution would skew lower due to the assumptions laid out in Sec.~\ref{sec:lpts} and thus the agreement is closer with respect to AM CVn-like systems.
}
 
\section{Orbital gravitational waves}  \label{sec:gws}

The base, dimensionless GW amplitude of a circular binary reads \citep{finn00}
\begin{equation} \label{eq:lisa0}
h \approx 2 \times 10^{-22} \left( \frac{1 \text{ hr}} {P} \right)^{2/3} \left( \frac {\mathcal{M}} {0.5 M_{\odot}} \right)^{5/3} \left( \frac {1 \text{ kpc}} {D} \right).
\end{equation}
From a data-analysis perspective, it is more convenient however to express the detectability via the dimensionless amplitude
\begin{equation} \label{eq:dimamp}
\begin{aligned}
    \mathcal{A} &= \frac{2 (G \mathcal{M})^{5/3} (\pi \fGW)^{2/3}}{c^4 D} \\
    &\approx 2 \times 10^{-22} \left( \frac{\fGW}{1 \text{ mHz}}\right)^{2/3} \left( \frac {\mathcal{M}} {0.5 M_{\odot}} \right)^{5/3} \left( \frac {1 \text{ kpc}} {D} \right).
    \end{aligned}
\end{equation}
Expression \eqref{eq:dimamp} is that which is typically used in the generation of synthetic catalogues \cite[e.g.,][]{korol22,cury22}. In this notation, the orthogonal components of the signal in the source frame read \citep{PhysRevD.76.083006}
\begin{equation} \label{eq:sourcecomponents}
    h_{+}^{\rm s}= \mathcal{A} \left(1 + \cos^2 \iota\right) \cos \left[\Phi(t)\right], h_{\times}^{\rm s}  = 2\mathcal{A} \cos \iota \sin \left[\Phi(t)\right],
\end{equation}
where
\begin{equation}
    \Phi(t) = \phi_0 + 2 \pi \fGW t + \pi \dot{f}_{\rm GW} t + \mathcal{O}(\ddot{f}_{\rm GW}).
\end{equation}

In total therefore, each synthetic catalogue we produce consists of an array of $8 \times N$ values pertaining to $\{\fgw,\fdgw,\mathcal{A},\lambda,\ell,\iota,\psi,\phi_{0}\}$ where the symbols are defined in the preceding sections. 
We now discuss how such arrays can be translated into SNRs before moving to the detectability results.

\subsection{Estimating the resolvable binaries and their residual confusion signal}  \label{sec:signaltonoise}

\begin{figure*}
\centering
 \includegraphics[width=0.8\textwidth]{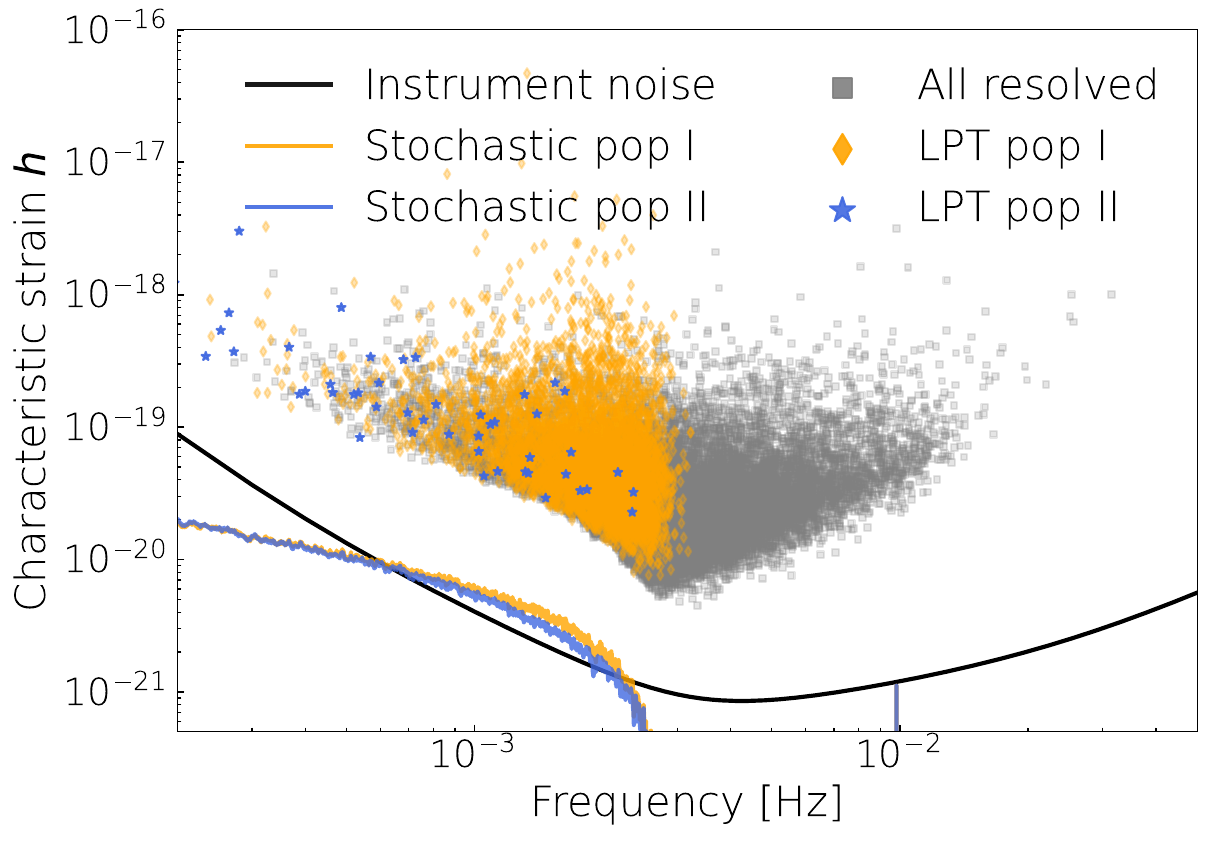}
 \caption{Characteristic strain, $h$, as a function of GW frequency, $\fGW$, for two synthetic LPT catalogues corresponding to Populations I (yellow diamonds) or II (blue stars). 
 Coloured points indicate an SNR of at least 7, relative to the instrument noise (black curve). 
 Overlaid in grey squares are resolvable WDWD binaries for context, generated using the specifications described in text.
 The coloured, solid curves correspond to confusion noises resulting from unresolved population of WDWD and LPT binaries; they exceed the instrument noise for $1 \lesssim \fGW/\text{mHz} \lesssim 2$.
 Strains are computed assuming a 4 year integration window. 
}
  \label{fig:gwstrain}
\end{figure*}

Using the LISA design specifications from \cite{LISA:2024hlh}, we adopt the noise power spectral density [PSD; $S_n(f)$] from \cite{Babak:2021mhe} which includes intrinsic noise (from test-mass acceleration and single-link optical metrology). 
We estimate the Galactic ``confusion'' noise from unresolved sources in order to assess LPT detectability. 
We employ the quasi-monochromatic wave-form described by eight parameters defined above in the array~\citep{PhysRevD.76.083006}, ignoring external and/or environmental factors that may introduce any distortion of this signal (e.g. data gaps, or other data non-stationarities). 
We then generate the waveforms for the catalogue entries and add them to the instrumental noise. 

In order to estimate the resolvable sources, as well as the stochastic component of their ensemble signal, we follow the iterative procedure and software described in~\cite{karn21}. 
At each iteration $i$ the data in the frequency domain are smoothened and adopted as the overall noise $S_{n,i}(f)$ (instrumental noise plus the stochastic GW signal). 
Then, the SNR of the sources is re-computed with respect to the new noise curve $S_{n,i}$, and any source that exceeds a certain SNR threshold ($\text{SNR}_\mathrm{thr} = 7$), is considered resolvable and then is subtracted from the data. 
The procedure repeats for iteration $i+1$, with $S_{n,i+1}$ as the new baseline for the overall noise. 
Convergence is reached when there are no more sources left to subtract, or when $S_{n,i}(f) - S_{n,i+1}(f) \leq \varepsilon$, for a given threshold of $\varepsilon$ for all frequencies $f$.
The chosen threshold matches that in the above reference.

The SNR of a given source is calculated through 
\begin{equation}
(\text{SNR}_\mathrm{tot})^2 = \sum_j  \left( h_j | h_j \right),
\label{eq:snrtot}
\end{equation}
with $j \in \{A, E, T \}$ the noise-orthogonal Time Delayed Interferometry (TDI) data channels of LISA~\citep{Tinto:2004wu}, 
and $\left( \cdot | \cdot \right)$ denotes the noise weighted inner product expressed for two time 
series $a$ and $b$, as
\begin{equation}
\left( a | b \right) = 2 \int\limits_0^\infty \mathrm{d}f \left[ \tilde{a}^\ast(f) \tilde{b}(f) + \tilde{a}(f) \tilde{b}^\ast(f) \right]/S_n(f).
\label{eq:ineerprod} 
\end{equation}

\section{Results} \label{sec:results}

\subsection{Resolvability} \label{sec:resolve}

Figure~\ref{fig:gwstrain} depicts the characteristic strain, $h$, for the sources produced in our two main catalogues for Populations I and II (Figs.~\ref{fig:fGWlist} and \ref{fig:chirplist}). 
To place these results in the broader context of Galactic WDWD binaries, which are expected to numerically dominate the LISA band, we also show a comparison sample \cite[e.g.][]{LISA:2024hlh}. 
The overlaid grey squares correspond to a separate catalogue from~\cite{Korol:2023jfz}, containing $\sim2\times10^{7}$ such systems overall.
Individual points, from a given catalogue, correspond only to those sources for which $\text{SNR} \geq 7$, as is often considered the marginal threshold for detection~\citep{karn21, 2023MNRAS.522.5358F, katz25}.
Such an SNR is normally computed relative to the \emph{instrumental} noise curve (shown in black); in reality, confusion noises from other sources also apply and influence the computation \citep[see, e.g.,][]{2025arXiv251018695P}.

\begin{figure*}
\centering
 \includegraphics[width=0.75\textwidth]{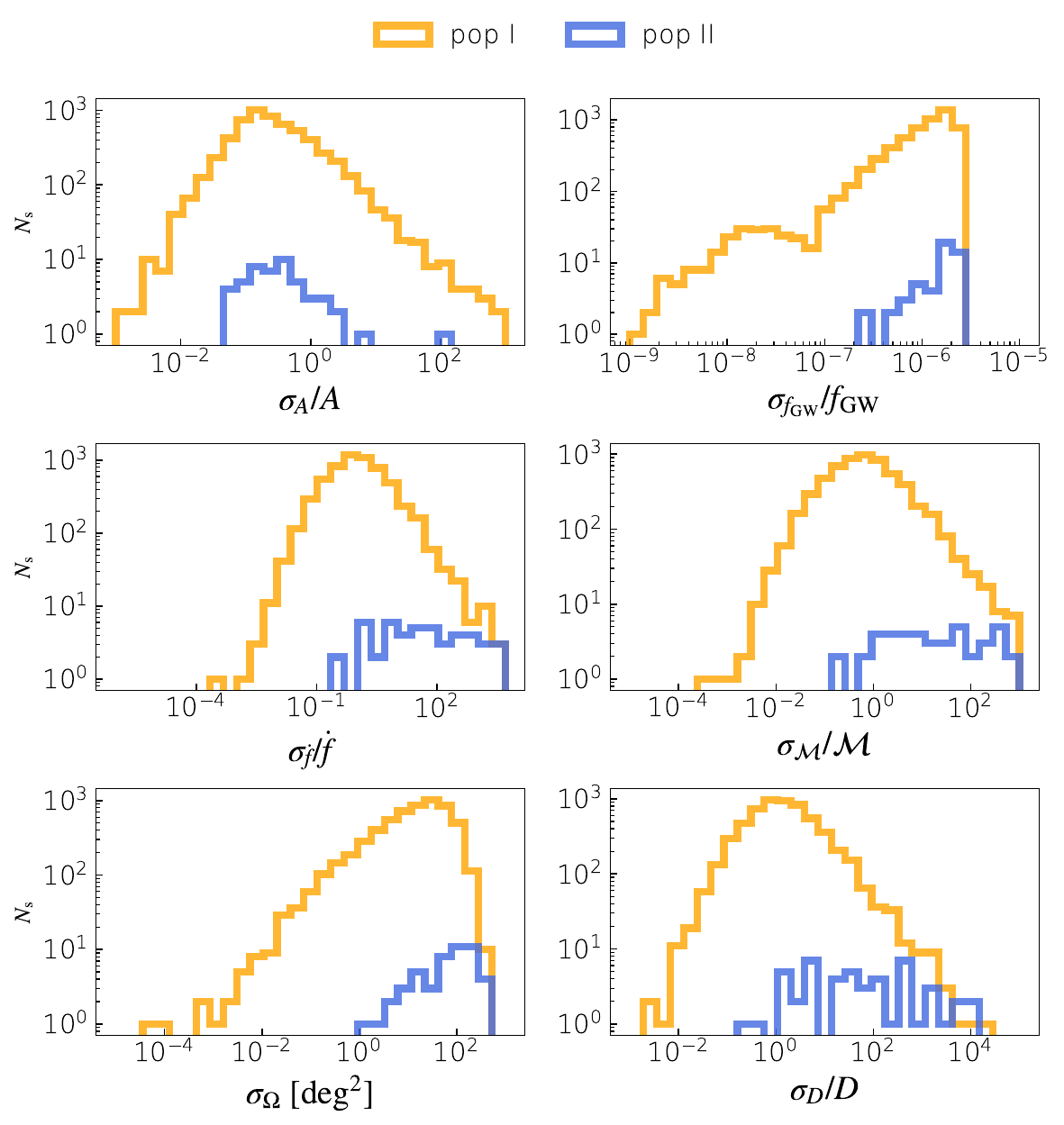}
 \caption{Results of a Fisher analysis for parameter estimation for LISA-resolvable sources ($\text{SNR} \geq 7$) from Populations I (yellow) and II (blue). The vertical axis shows the raw count from the synthetic catalogues, with the horizontals showing the fractional recoverabilities ($\sigma_{\lambda^{i}}/\lambda^{i}$), at the $68 \%$ confidence level, for parameters $\lambda^{i}$ being: the GW amplitude (top left), frequency (top right), frequency derivative (middle left), chirp mass (middle right), and luminosity distance (bottom right). The bottom left panel shows instead the sky localisability ($\sigma_{\Omega}$) in units of square degrees.}
  \label{fig:fisher}
\end{figure*}

Points shown with yellow diamonds correspond to those in Population I: we see a litter of resolvable sources within the LISA band up to $\sim 4$~mHz, corresponding to the value of $P_{L}$ selected in the CDF in Eq.~ \eqref{eq:periodcdfA}. 
{In total we estimate that $\approx 5911$ of Population I systems would be (at least marginally) detectable, corresponding to a fraction $\sim 0.06$ (i.e., $6\%$).}
As expected, this case is relatively optimistic: many of the known LPTs show sub-hour pulse periods (Table~\ref{tab:lptdata}), and thus the empirical construction with $\alpha = 1/2$ leads to a pileup of resolvable systems at $\gtrsim$~mHz frequencies.
{At these frequencies we expect degenerate companions, and thus the Roche-lobe overflow condition \eqref{eq:radiuscond} is easier to fulfill which allows for generally-larger chirp mases.}
If we were to demand instead $\text{SNR} > 14$ (i.e., twice the floor value used to select points), this fraction drops to {$ 0.017$ ($0.00013$) for Population I (II)}; within the catalogue, we find recoverable sources with SNRs as high {as $\sim 446.5$ (the single source with dimensionless strain $h \approx 4 \times 10^{-17}$ around $\fGW \approx 1.5$~mHz)}.
Either way, this indicates that a non-negligible portion of the {approximately phase-locked} LPT {subpopulation} could be detectable by LISA and thus such systems may be multimessenger (see Sec.~\ref{sec:outlook}).
Indeed, without direct radio or other observations it could be unclear whether a detected source is an LPT or some other type of binary.

In contrast to the above, points shown with blue stars correspond instead to those from Population II.
{This case is much more pessimistic: we estimate that only $\sim 49$ sources are resolvable, corresponding to a small fraction $\sim 5 \times 10^{-4}$ of the total.
The highest SNR found is $\approx 131.5$ -- much lower than the brightest Pop I source.}
This is again expected as $\alpha = 10/3$ skews the frequency distribution towards lower values (Fig.~\ref{fig:fGWlist}).

The estimated confusion foreground arising from the Galactic population of WDWD binaries and LPT populations is shown as yellow and blue solid curves. 
We remind the reader that this is an incoherent superposition of many unresolved signals that will contribute additional noise for all (extra-)galactic LISA sources.
Both curves rise above the instrument noise between 0.7 and 3 mHz. 
This is primarily due to the WDWD binary population. 
We note however that the induced confusion noise including Pop II LPTs (blue curve) is somewhat weaker. 

As a consequence of the milder constraints on the onset of accretion (Eq.~\ref{eq:radiuscond}), Pop II systems remain in the LPT phase for longer, leading to a higher density of detectable sources at sub-mHz frequencies compared to Pop I. 
Given a sufficient number of detections, the distribution of source frequencies can be used to infer the underlying orbital evolution, and thus constrain the coefficient $\alpha$. 
A value of $\alpha \approx 7/3$ would point to GW-dominated evolution while $\alpha \approx 10/3$ indicates unipolar-inductor-driven decay, for example.
Beyond confirming the binary nature of LPTs, LISA observations can therefore provide insight into their formation channels and emission physics (e.g., by observationally comparing electromotive decay power to the radio luminosity to deduce the efficiency $\epsilon$). 

\subsection{Parameter estimation} \label{sec:parameterrecovery}

We now turn to the related question of how well system parameters may be recovered given a hypothetical future detection of an LPT-like system.
To achieve this, we use a Fisher-matrix analysis\footnote{Such a scheme is strictly only valid at very high SNRs \protect\citep{fish35}. However, at the level of the analysis carried out in this paper it provides a reasonable approximation. A more thorough investigation, involving convolutions and a Bayesian analysis, will be considered in future work.}, mathematical details of which are deferred to Appendix~\ref{sec:fisher} in an effort to not interrupt the flow of text.

The results for anticipated recovery errors (at a $68\%$ confidence level) in individual parameters are shown in Figure~\ref{fig:fisher} in terms of absolute source count ($N_{\rm s}$).
In general, as a greater number of sources are detectable from Pop I (yellow) over Pop II (blue), the values of $N_{\rm s}$ are greater for each injected parameter in any given bin in the former case.

\begin{figure*}
\centering
 \includegraphics[width=0.487\textwidth]{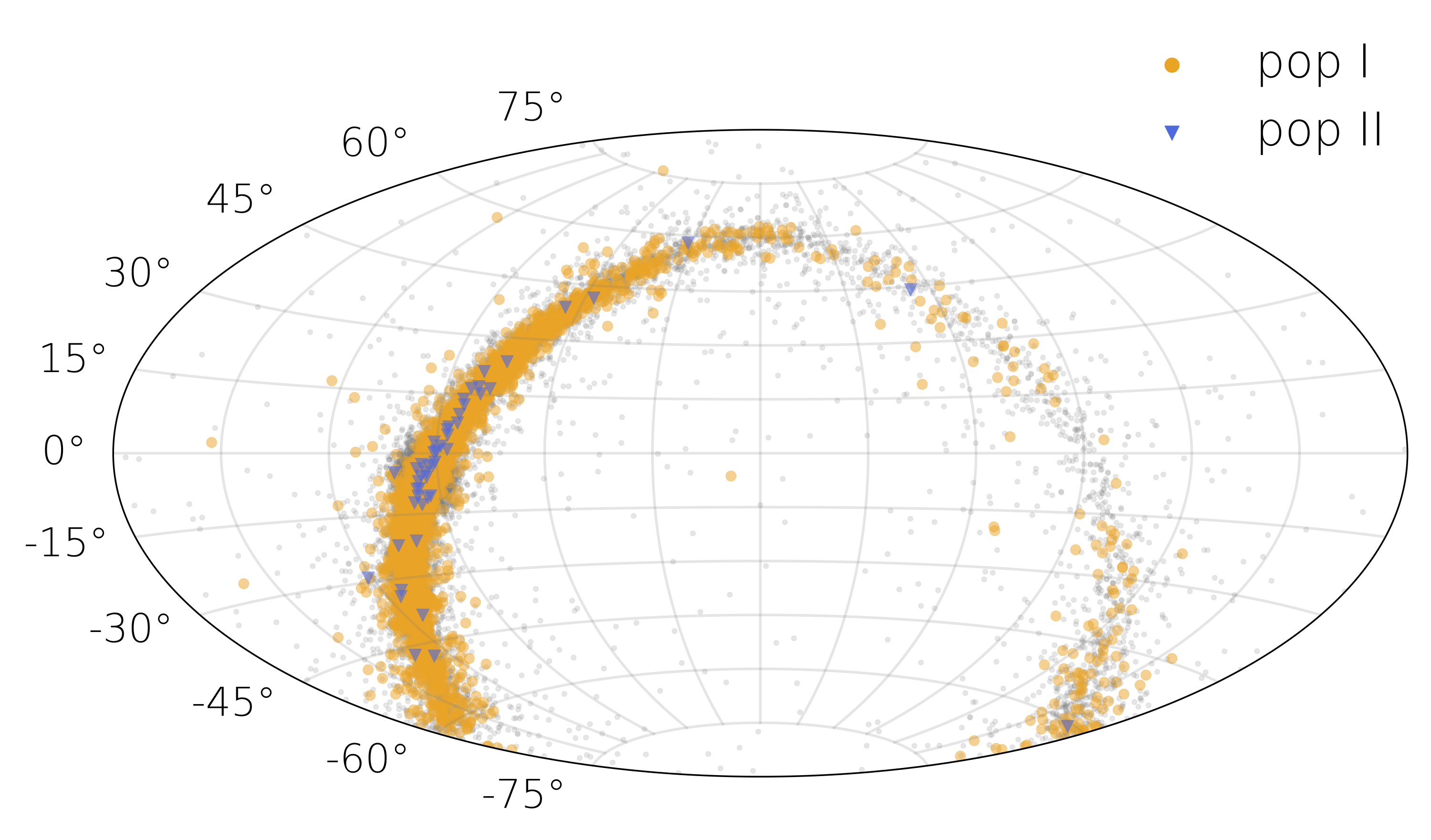}
 \includegraphics[width=0.487\textwidth]{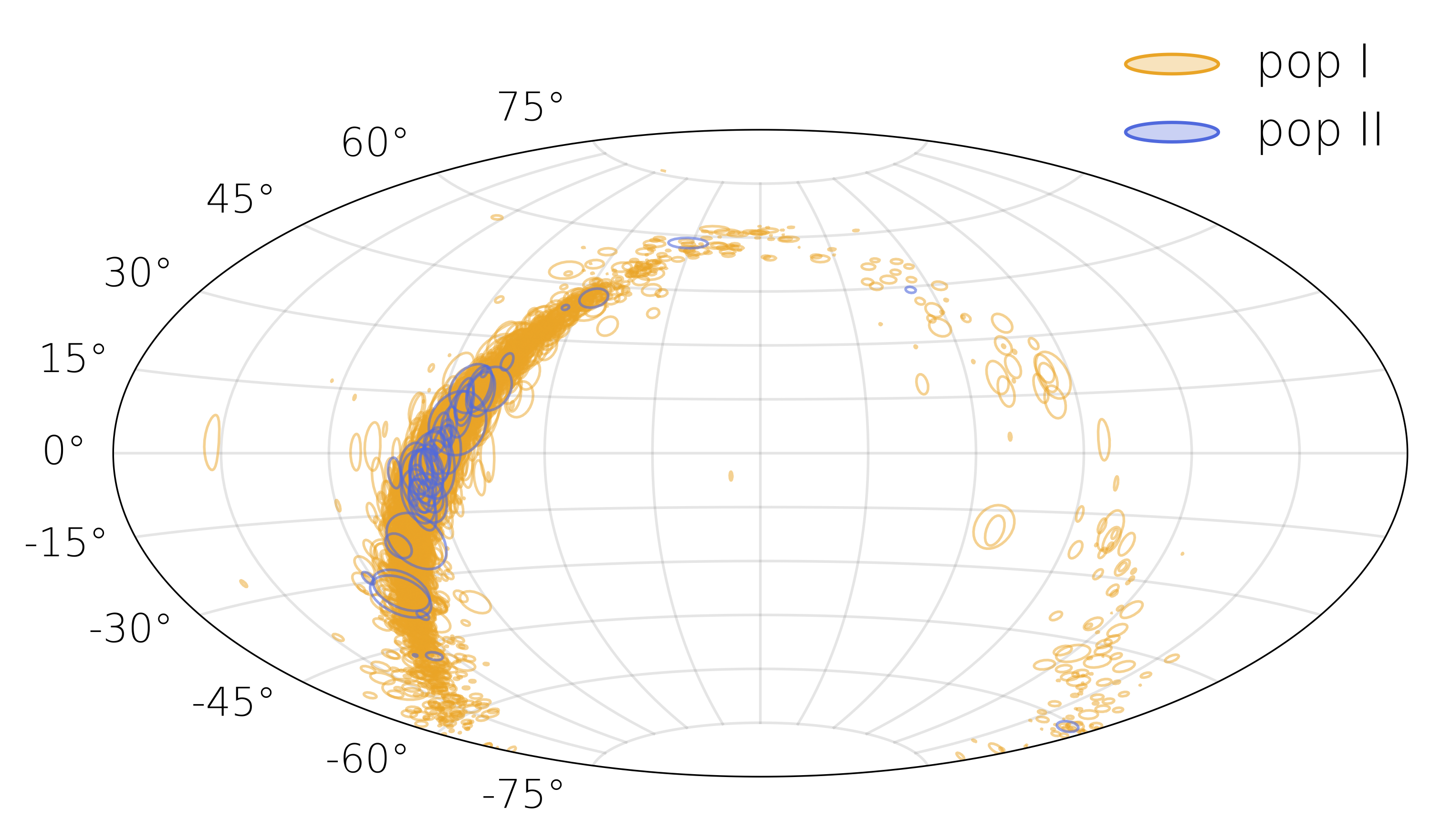}
 \caption{Left: Distribution of the resolved LPTs for the two populations considered. 
 Recovered binaries of pop I are represented with yellow circles, while pop II are shown with blue triangles. 
 For contrast, we show with gray points expectations for the distribution of recoverable WDWD binaries. 
 Right: Similar to the left panel, though including the localization errors for each recovered (LPT) source.}
  \label{fig:sky}
\end{figure*}

The top panel shows the retrievability of the amplitude ($\mathcal{A}$; left) and the GW frequency ($\fGW$; right). 
For the amplitude, we see that the bulk of the sources have $\sigma_{\rm A}/\mathcal{A} \leq 1$ for either population, implying a typical error of less than a factor $\sim 2$. 
As is clear from the right-hand panel however, the frequency can be recovered exceptionally well --- {to at worst one part in $\sim 10^{5}$ in almost all cases}.
If the source distance is known (cf. Fig.~\ref{fig:distperiod}), this means that amplitude recoverability can be directly translated into constraints on the chirp mass via Eq.~\eqref{eq:dimamp}.
For example, we see that the top-left histogram peaks at values of $\sigma_{\rm A}/\mathcal{A} \approx 10^{-1}$. 
This equates to a $\approx 6\%$ absolute error bar on $\mathcal{M}$ if other parameters are exactly known. 
This would likely be sufficient to resolve the individual component masses given $\fGW$ information by making use of Kepler's law, highlighting again the value of targeted LISA searches for LPTs (see Sec.~\ref{sec:outlook}). 

On the other hand, if we do not assume any prior knowledge on the luminosity distance, LISA data alone may not be able to provide strong constraints on $\mathcal{M}$ itself even given tight $\mathcal{A}$ and $\fGW$ values. 
Significant errors may be anticipated on the distance in general (as many sources have $\sigma_{\rm D}/D \gg 1$; bottom right panel).
This means that a blind, independent LISA search leads to weaker constraints one could place on $\mathcal{M}$ and hence binary nature (middle right panel). 
This is not only due to distance uncertainties but also from the frequency derivative (middle left panel), where again we have many cases where $\sigma_{\dot{f}}/\dot{f} \gg 1$: a loose constraint on $\dot{f}_{\rm GW}$ implies a weaker measurement of $\dot{E}$ (cf. Eq.~\ref{eq:pdotgw}) and thus $\mathcal{M}$.
Even without multimessenger input, however, LISA data could prove useful to rule out some progenitors \cite[such as isolated neutron stars, which are still viable for a number of the known systems;][]{suvm23,ben23,coop24,lyman25}. 

Aside from tail events, the general conclusions reached above apply broadly to both populations.
That is, both sets of distributions follow roughly the same patterns (e.g., peaking at similar values for each $\sigma_{\lambda^{i}} / \lambda^{i}$).
One major difference, however, is the absolute source count: there is a preferential tail for high SNR events in Pop I, with the brightest sources allowing for extremely tight constraints to be placed. 
{We see that there are $\sim 20$ sources ($\sim 0.02\%$ of the total) for which we could expect $\mathcal{M}$ and $\dot{f}_{\rm GW}$ to be constrained to better than one part in $\sim 10^{2}$.}
A resolution of this quality could be used to conclusively deduce the nature of orbital decay and hence even identify the ratio $\dot{E}_{\rm GW}/\dot{E}_{\rm UI}$ in addition to the component masses.
Such constraints could guide radio surveys for LPTs, as a value of $\epsilon$ translates into a radio luminosity given $\dot{f}_{\rm GW}$ (see, e.g., Eqs.~\ref{eq:fdotuiILT} and \ref{eq:fdotuiGX}).

Another important parameter we can extract from the Fisher analysis relates to the sky localisability in terms of solid angle ($\Omega$; bottom left panel). 
Here, the values of $\sigma_{\Omega}$ are shown in square degrees, where we see the distribution {peaking at values of $\sim 50 \text{ deg}^{2}$ for population I, while results are more pessimistic for population II. }
How source \emph{position} can be used to guide electromagnetic surveys is explored in the following.

\subsection{Multimessenger outlook for the cohort of binary LPTs} \label{sec:outlook}

\begin{table*}
\caption{Comparison of the observing frequencies and field-of-views for a variety of radio telescopes, both planned and existing, relevant for the discovery of LPTs. 
Most data listed here are reproduced directly from Table 3 in \protect\cite{hassall13} and references therein. Following this reference, FOV values are estimated using the centre of the quoted observing band. 
Some instrument names have been abbreviated.}
\begin{center}
\begin{tabular}{lccccc}
\hline
Telescope & $\nu_\mathrm{low}$ & $\nu_\mathrm{high}$ &  FOV & Reference(s)\\
		 & (MHz)	& (MHz) & (deg$^2$) \\
         \hline
\hline
SKA-low  & 50 & 350 & $\sim$27 & \cite{dewd13}\\
SKA-mid  & 1000 & 2000 & $\sim$0.5 & \cite{dewd13}\\
LOFAR-HBA  & 155 & 165 & $\sim$150 &\cite{stap11} \\
LOFAR-LBA  & 30 & 80 & $\sim$100 & \cite{stap11} \\
MWA  & 185 & 215 & $\sim$375 & \cite{ting13} \\
ASKAP  & 700 & 1000 & $\sim$30 & \cite{john09}\\
MeerKAT  & 580 & 1750 & $\sim$1.0 & \cite{blok10} \\
Parkes Multi-beam  & 1230 & 1518 & $\sim$1.1 & \cite{manchester01}\\
VLA & 350 & 1400 & $\sim$1 &  \cite{cond98,betha21}\\
CHIME & 200 & 800 & $\sim$200 &  \cite{chime18}\\
GBT & 290 & 1400 & $\gtrsim$0.3 &  \cite{pilla21}\\
\hline
\hline
\end{tabular}
\end{center}
\label{tab:radiotel}
\end{table*}

The first question to ask is independent of LISA: given a Galactic population of LPTs, how many might one expect to be visible as radio sources? 
Owing to their long periods, LPTs are difficult to identify within traditional surveys and were only discovered once `fast imaging' techniques were used \cite[see section 3.9 in][for a discussion]{murph26}. 
Moreover, given that the anticipated time-of-arrival noise decreases only as the square root of the pulse count \citep{lk12}, obtaining a reliable measure of $P$ generally requires many pulses to be recorded (even more so for $\dot{P}$). 
If, however, a source frequency (see Fig.~\ref{fig:fisher}) is detected by LISA, the recording of even two pulses should be sufficient to confirm the source as an LPT and not some other type of Galactic binary or radio transient. 
With this in mind, the question posed above has been considered in detail by \cite{horv25} and \cite{lee25b}, focussing in particular on the GLEAM-X and EMU surveys as part of the Murchison Widefield Array (MWA) and Australian Square Kilometer Array Pathfinder (ASKAP), respectively \cite[see also][]{murph26}.
For example, Appendix A of the former reference estimates that sources with periods of $\sim 10$~min would have a $> 95\%$ probability of having at least 2 pulses detected if within the FOV of the MWA. 
For hypothetical sources with a period of a few hours, this chance drops to $\approx 40\%$.

The above estimates show that multimessenger-followup is promising {in optimistic cases}. 
Given the overlap between the GLEAM-X data-release sky maps \cite[DRI and DRII; see figure 2 in][]{horv25} and the distribution considered in Sec.~\ref{sec:populationnumber} together with the wide FOV of the MWA, we estimate that more than $\sim 80\%$ of ``short-period'' LPTs could be identified in a future GLEAM-type survey using the solid angle localisation provided by a LISA detection. 
To facilitate a direct comparison, we show in Figure~\ref{fig:sky} the sky-localisation errors predicted for each of the sources identified as detectable in Pops I and II.
We list in Table~\ref{tab:radiotel} relevant FOVs for a number of radio telescopes that have detected LPTs in addition to planned constructions. 
Many have an FOV that is sufficiently wide that the localisation that could be achieved with LISA would easily be enough to identify the sky location in which one may scan for pulsations. 
Others, however, are more borderline. 
For example, in the $1$--$2$~GHz band with SKA-mid the design FOV is $\sim 0.5$~square degrees \citep{dewd13}.
{Comparing directly with the bottom-left panel of Fig.~\ref{fig:fisher}, we see that a fraction $\sim 3\%$ of LISA-detected sources with $\text{SNR} \geq 7$ would be localised well enough to have pulsations searched for with SKA-mid for Population I.
For Population II, the best-resolved source already is such that $\sigma_{\rm \Omega} \approx 1 \text{ deg}^2$.}

\begin{figure}
\centering
 \includegraphics[width=0.487\textwidth]{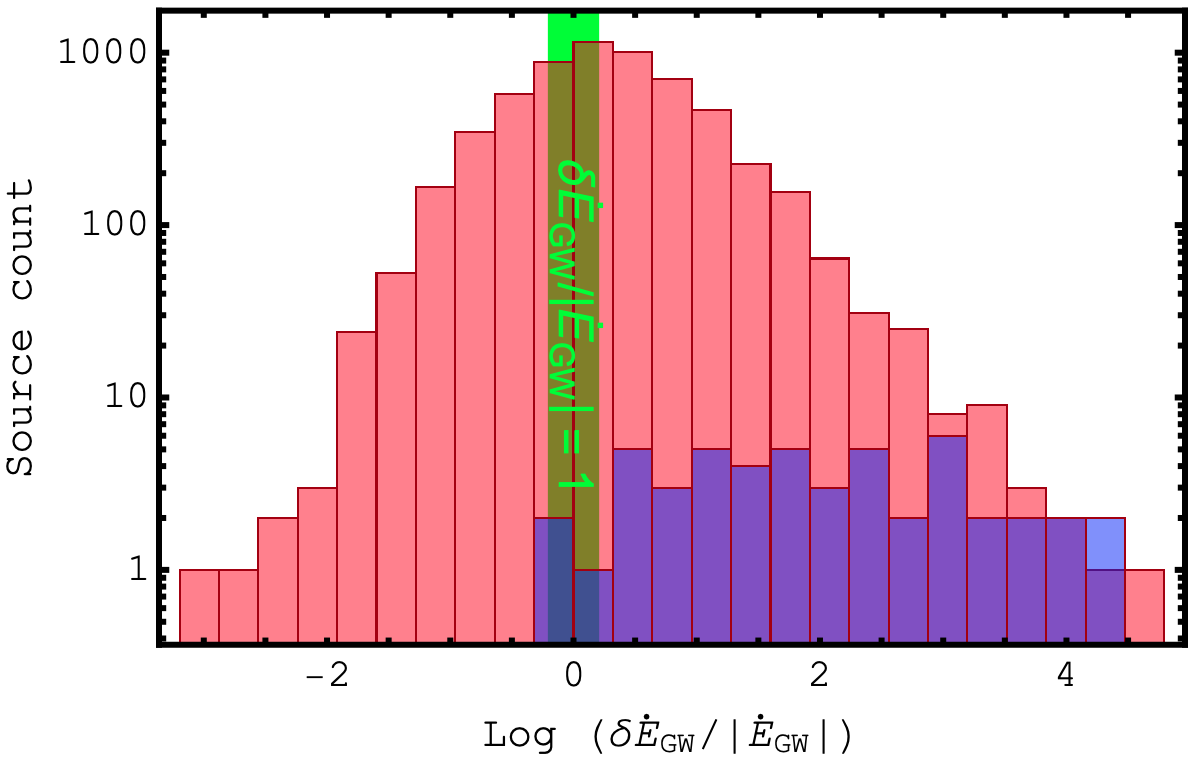}
 \caption{Fractional uncertainty on the energy-decay rate \eqref{eq:uncert} given uncertain measurements of $\fGW$, $\mathcal{M}$, and $\dot{f}_{\rm GW}$. Population I sources with $\text{SNR} > 7$ are shown in red, with Pop II in blue. 
 The green, vertical bar depicts unity, ${\delta \dot{E}_{\rm GW}}/{|\dot{E}_{\rm GW}}| = 1$; sources to the left of this box should allow for an inference of whether electromotive losses are active.}
  \label{fig:uncerts}
\end{figure}

Given measurements of a triplet $\{\mathcal{M}, \fGW, \dot{f}_{\rm GW}\}$, a comparison with the GW-decay prediction from Eq.~\eqref{eq:pdotgw} can be made. 
In particular, one may anticipate a mismatch between the observed and GW-predicted values if electromotive losses are significant \citep{lai12,skk24}. 
If the orbit were decaying \emph{faster} than that predicted by Eq.~\eqref{eq:pdotgw}, this would point strongly to excess losses which could come to fuel radio emissions. 
On the other hand, if $\dot{P}$ is lower (or even positive) then there may be accretion torques stalling the inspiral. 
To provide some estimates in this respect, note that Eq. \eqref{eq:orbdecayeqn} implies $\dot{E}_{\rm GW} \propto \mathcal{M}^{5/3} \fGW^{-1/3} \dot{f}$. 
If we assume independent measurements of $\fGW$, $\dot{f}$, and $\mathcal{M}$, and posit some uncertainties ($\delta$), we can estimate the fractional uncertainty
\begin{equation} \label{eq:uncert}
    \frac{\delta \dot{E}_{\rm GW}}{|\dot{E}_{\rm GW}|} \approx \sqrt{\left(\frac{1}{3}\frac{\delta f_{\rm GW}}{f_{\rm GW}}\right)^2 + \left(\frac{\delta \dot{f}}{\dot{f}}\right)^2 +\left(\frac{5}{3}\frac{\delta \mathcal{M}}{\mathcal{M}}\right)^2} \, .
\end{equation}
The first term here is negligible in cases of interest, though the latter two could be large for low SNR. 
In particular, from Eqs.~\eqref{eq:fdotuiILT} and \eqref{eq:fdotuiGX}, we expect an unambiguous detection of electromotive losses could be achieved if ${\delta \dot{E}_{\rm GW}}/{|\dot{E}_{\rm GW}}| \ll 1$. 
Figure~\ref{fig:uncerts} depicts relation \eqref{eq:uncert} from the Fisher-matrix data.
{We see that, for the typical parameter-recovery errors shown in Fig.~\ref{fig:fisher}, we expect ${\delta \dot{E}_{\rm GW}}/{|\dot{E}_{\rm GW}}| \lesssim 1$ (solid contour) in $\approx 35\%$ of cases for Population I.
For Population II, however, it is likely that the mechanism responsible for orbital decay would be difficult to identify from LISA data alone as only $\approx 4\%$ are such that ${\delta \dot{E}_{\rm GW}}/{|\dot{E}_{\rm GW}}| < 1$.
Either way, positional constraints can be cross-referenced with optical catalogues like Pan-STARRS or X-ray data to provide independent confirmation or otherwise \citep[see, e.g.,][]{rod25,ak25}. }

\begin{figure*}
\centering
 \includegraphics[width=0.9\textwidth]{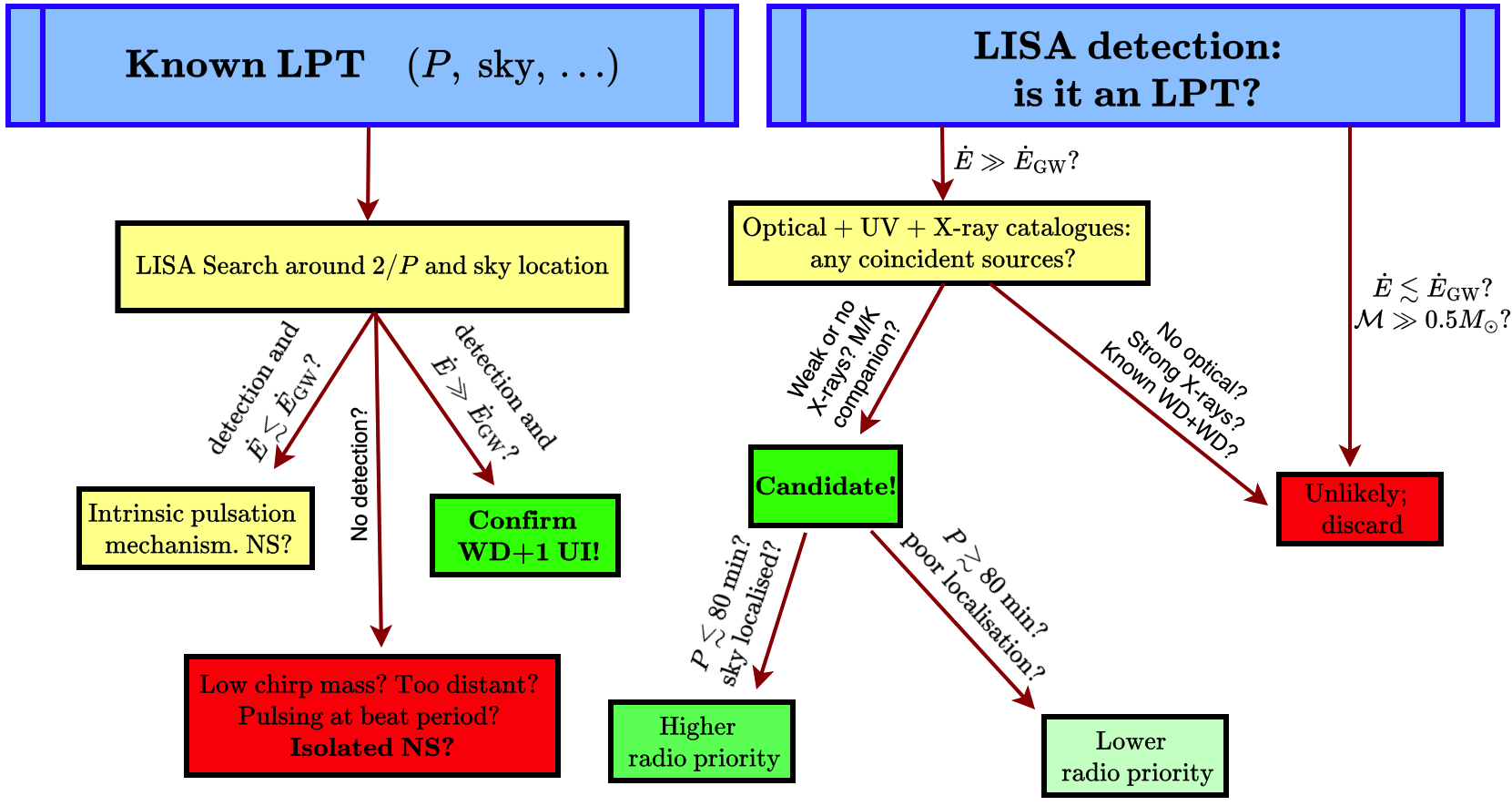}
 \caption{Flowchart for strategies regarding multimessenger followup starting from either a known LPT (left branches) or a LISA detection (right). In the left tree we assume the period and other data (e.g., sky position) of an LPT is known. Depending on the results of a LISA search, one may be able to identify whether a unipolar inductor (UI) is active in a WD system with a companion (WD+1). If the search fails to detect the source, this may indicate that that particular LPT is not a binary but rather an isolated neutron star (NS), for example. The right panel instead traces a decision tree where a compact binary is detected and one wishes to assess the likelihood of it being an LPT by performing several checks.}
  \label{fig:flow}
\end{figure*}

In summary, we expect that those sources which are (1) X-ray dim, (2) optically bright, and (3) with $\dot{E} \gg E_{\rm GW}$ are the optimal candidates to target for followup. 
A flowchart, depicting a possible set of multimessenger strategies, is shown in Figure~\ref{fig:flow}.
On the other hand, a number of LPTs show low radio duty cycles \citep{rea26} and it may be difficult to identify \emph{which} LISA sources are the best candidates for followup with radio instruments.
In other words, the source could be dormant even if it sometimes manifests as an LPT.
In the pre-CV model, \cite{yang25} estimates typical duty cycles of $10^{-3} \lesssim d_{\rm radio} \lesssim 10^{-1}$: one may therefore have to monitor the relevant sky patch for a durations up to $T_{\rm obs} \sim 10^{3} P \approx 40 (P/1\text{ hr})$~days to hunt for a LISA-detected LPT.

\section{Discussion and conclusions} \label{sec:conclusions}

LPTs are mysterious, radio-pulsing sources. 
Some studies favour neutron-star progenitors while others stipulate compact binaries involving WDs \cite[see][for a sample]{suvm23,wang24,lan26}. 
The latter position is arguably more defensible given that some LPTs are now confirmed via optical spectroscopy and broadband photometry to contain type M companions \citep{hw24,rui24}. 
If radio pulsations are phase-locked to the orbit—as is the case for the confirmed binaries cited above—some fraction of the Galactic cohort \cite[of order $N \sim 10^{5}$ or more;][]{yang25,rea26} of ``short-period'' LPTs could be visible to LISA.
If electromagnetic data remain inconclusive as more sources are discovered, GW observations could constrain their nature. 
By generating two source catalogues, based on either extrapolating from the known sources (Tab.~\ref{tab:lptdata}; Pop I) or assuming electromotive losses dominate the orbital evolution (Pop II), {we found that between $\sim 6\%$ and $\sim 0.05\%$ of LPTs should be visible to LISA, respectively.}
The main results are captured by Figs.~\ref{fig:gwstrain} and \ref{fig:fisher}.

The first of these shows the characteristic strains, as a function of frequency, for the sources injected into a LISA data-analysis pipeline following \cite{karn21} and others, for which the SNR is greater than the detection threshold (see Sec.~\ref{sec:signaltonoise}).
{For Pop I sources we estimate that $\sim 6\%$ of the Galactic LPT population could be detectable ($\sim 6000$ sources), which drops to $\sim 0.05\%$ for those in Pop II ($\sim 50$ sources).}
These two cases were chosen to represent the plausible extrema of how the periods may be distributed.
The true value is likely between them because there may be subpopulations of LPTs, depending not only on age and the evolutionary stage \cite[e.g.,][]{yang25} but progenitor type \cite[e.g.,][]{afon24}.

The second Figure shows instead how well source parameters can be recovered: even if the count ($N_{s}$) is low, the findings are optimistic in the sense that the amplitudes (top left), GW frequencies (top right), and sky locations (bottom left) should be well-constrained following even a marginal detection.
Importantly, we find the histogram of recovered $\sigma_{\fGW}/\fGW$ values peaks at $\gtrsim 10^{-6}$.
Given this may be twice the value of the reciprocal of the pulse period (Eq.~\ref{eq:gwfreq}), or close to depending on { spin slippage}, such a strong constraint could help guide data-reduction pipelines which search for radio pulsations with high-FOV instruments (Tab.~\ref{tab:radiotel}) as the patch of sky is typically constrained to within $\sim 30 \text{ deg}^{2}$ ({i.e., the median value;} see also Fig.~\ref{fig:sky}).
This means that not only can LISA be used to constrain the nature of known LPTs \citep{suv25}, but that it could help to find new candidates (as discussed in detail in Sec.~\ref{sec:outlook}).

In the phase-locked, pre-CV picture adopted here, we expect that the companion and the primary are connected through a magnetospheric flux tube which is effectively stationary (Fig.~\ref{fig:schematic}).
Particles accelerating along the tube eventually slam into the WD atmosphere, potentially creating a persistent auroral hot spot. 
The spot should not move across the stellar surface; rather, the star, the companion, and the spot will rotate rigidly and we expect consistent but low-flux emissions. 
We may thus see a pulse of X-ray emission every time the spot rotates into a detector line-of-sight if the dissipation occurring near the magnetic anchor points is strong enough (see Fig.~\ref{fig:schematic}).
Such a picture is consistent with the persistent X-ray detections in the LPTs ASKAP J144834--685644, J1912--4410, {and ASKAP J174508.9--505149}  at levels of $\sim 10^{-14} \text{ erg cm}^{-2} \text{ s}^{-1}$ \citep{ak25}, $\gtrsim 10^{-13} \text{ erg cm}^{-2} \text{ s}^{-1}$ \citep{peli23}, {and $\sim 3 \times 10^{-13} \text{ erg cm}^{-2} \text{ s}^{-1}$ \citep{imbrog26}} respectively.
By contrast, a \emph{burst-like} detection of X-rays in an LPT would disfavour the model, as an isolated neutron star or actively accreting WD becomes a more appealing progenitor \citep{suvm23,ben23,wang24}. 

With respect to the model presented here, there are a number of simplifying assumptions we have employed.
One relates to the orbital dynamics: we assumed circular orbits throughout.
For a source with modest eccentricity, $e$, the total signal will comprise a weaker (in terms of power) component at $\fGW$ together with multiple, narrow-band harmonics at frequencies of $\approx n \fGW$ for integer $n$ together with precessional nodes \citep{pm63}.
The power distribution into these harmonics depends on a complicated function representable as a sum of Bessels; see \cite{hamers21} for the exact expression and some practical fittings.
Using formulae therein, we estimate that for $e \approx 0.1$ the circular-component power reduces by $\approx 5\%$ while the next-strongest harmonics contain $\approx 11\%$ of the base power. 
Since the SNR scales as the square root of the power, such sidelobes will be undetectable unless the SNR is very high.
It is likely therefore that moderate eccentricity can only invite confusion noises and reduce the visibility of a given LPT \cite[see also][]{moore24}.
At present, however, there is no evidence for eccentricity in the known sources, though time will tell as we approach LISA launch and more LPTs are uncovered in radio.

For the individual LPT ASKAP J144834--685644, \cite{ak25} favour a nearly edge-on binary with a WD primary. 
There is an interesting possibility therefore that emitted GWs will be periodically \emph{lensed} as they are eclipsed by the companion once per orbit.
Measuring the time delay from an eclipse,
 $   \Delta \tau \propto \Mc \log\left( {2a}/{R_{\rm c}} \right)$,
combined with $\Mc$ from $\mathcal{M}$ and $P$, would provide detailed information on the hydrostatic structure of the companion even without optical data (cf. Eq.~\ref{eq:mdmrreln}), and thus further map out LPT genealogy.
One could compare the retrieved inclination, $\iota$, from the recovery pipeline to devise a subroutine to assess if self-lensing is likely to occur in any given source. 

\begin{acknowledgements}
{We thank the anonymous reviewer for providing critical feedback on evolutionary pathways, which improved the quality of this work.}
AGS thanks Marta Piscitelli for invaluable assistance with Python packages, Elias Most for clarifications regarding cyclotron masers, and the Center of Astrophysics and Gravitation (CENTRA) in Lisbon together with the Center of Gravity (CoG) within the Niels Bohr Institute in Copenhagen for hospitality shown while this work was being completed. 
This paper is part of a project that has received funding from the European Union's Horizon Europe Research and innovation programme under Grant Agreement No 101131928, project ACME. 
AGS acknowledges funding from the European Union's Horizon MSCA-2022 research and innovation programme ``EinsteinWaves'' under grant agreement No. 101131233 and the Deutsche Forschungsgemeinschaft through individual research grant 570901071. 
NK was supported by the Hellenic Foundation for Research and Innovation (H.F.R.I.) under the 4th Call for HFRI Research Projects to support Post-doctoral Researchers (Project Number: 28418).
\end{acknowledgements}

\appendix

\section{Broken power-law populations} \label{sec:brokenpower}

{In this Appendix, we consider a distribution of orbital periods -- arguably more realistic than that considered in Sec.~\ref{sec:sample} -- where instead a broken power-law CDF of the form
\begin{equation}
\tilde{F}(P) = \frac{1}{\mathcal{N}} \begin{cases} 0 & P \le P_{L} \\ P^{\alpha_1} - P_{L}^{\alpha_1} & P_{L} < P < P_{c} \\ P_{c}^{\alpha_1} - P_{L}^{\alpha_1} + \frac{\alpha_1}{\alpha_2} P_{c}^{\alpha_1 - \alpha_2} \left( P^{\alpha_2} - P_{c}^{\alpha_2} \right) & P_{c} \le P \le P_{U} , \end{cases}
\end{equation}
is used to set the base distribution, generalising expression \eqref{eq:periodcdf}, where 
\begin{equation}
\mathcal{N} = \left(P_{c}^{\alpha_1} - P_{L}^{\alpha_1}\right) + \frac{\alpha_1}{\alpha_2} P_{c}^{\alpha_1 - \alpha_2} \left( P_{U}^{\alpha_2} - P_{c}^{\alpha_2} \right)
\end{equation}
is a normalisation constant set to ensure that $\tilde{F}(P_U) = 1$. 
In the above, we have two arbitrary power-law indices, $\alpha_1$ and $\alpha_2$, and a cutoff period, $P_{c}$.
The latter can be set as $80$~min to delimit the ``pre-polar'' and AM CVn-like populations where the mass-transfer timescale becomes shorter than the thermal Kelvin-Helmholtz timescale.
Furthermore, we suppose that $\alpha_2 = 10/3$ for UI torques to apply to high-period sources and $\alpha_1 = 7/3$ so that GW losses dominate the lower track, in accord with theoretical expectations for highly-magnetised WD primaries \cite[see, e.g.,][]{nele01,nele04}.
A tapering, similar to that in equation \eqref{eq:periodcdfA}, is then applied to the high-frequency sources,
\begin{equation} \label{eq:broken}
F(P) = \begin{cases} 0 & P \le P_{L} \\ \tilde{F}(P) \exp\left[ \frac{s}{(P_{c} - P_{L})^2} - \frac{s}{(P - P_{L})^2} \right] & P_{L} < P < P_{c} \\ \tilde{F}(P) & P_{c} <P \leq P_{U}, \end{cases}
\end{equation}
again to prevent an unphysical pile-up at periods $P \gtrsim P_L$.
Because $F(P)$ is continuous everywhere and strictly increasing from 0 to 1, it serves as a valid CDF.
Such a choice represents a generalisation of the Population II model (and agrees in the limit $\alpha_1 = \alpha_2 = 10/3$).}

\begin{figure}
\centering
 \includegraphics[width=0.487\textwidth]{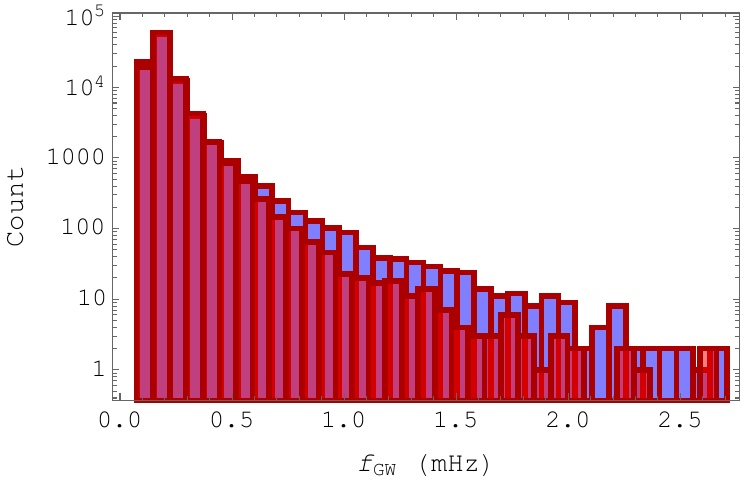}
  \includegraphics[width=0.487\textwidth]{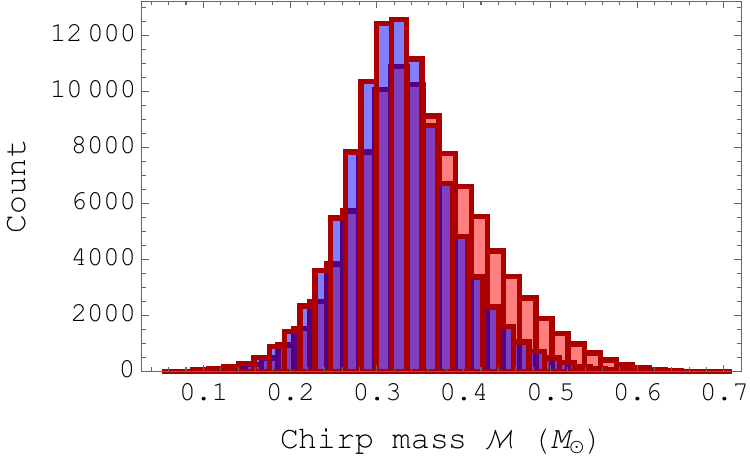}
 \caption{Similar to Figs.~\ref{fig:fGWlist} and \ref{fig:chirplist} but showing GW frequencies (top panel) and chirp masses (bottom) for a broken power-law drawing via expression \eqref{eq:broken} with $\alpha_2 = 10/3$, $\alpha_1 = 7/3$, and $P_c = 80$~min. The blue population corresponds to the broken power-law, while the red is that of Population II from the main text.}.
  \label{fig:fchirp3}
\end{figure}

{The results of drawing $N=10^5$ periods and subsequently chirp masses, following the methodology described in Sec.~\ref{sec:chirpmasses}, are shown in Figure~\ref{fig:fchirp3}.
Overall, as only $\lesssim 3\%$ of sources have periods below $80$~min thanks to the high value of $\alpha_2$, the distributions are largely similar to that presented in the main text for Population II.
A key difference, however, is visible in the GW frequencies (top panel), where the lower value of $\alpha_1$ relative to $\alpha_2$ produces a tail at higher frequencies; for instance, $0.38\%$ of sources have $f_{\rm GW} > 1$~mHz in this particular drawing.
The chirp distributions (bottom panel) are, by contrast, more similar because high-period sources dominate the count, though the values are skewed slightly lower as higher frequencies demand more compact sources (see Sec.~\ref{sec:lpts}); note also the linear scale of the histogram in this case, where sampling effects can adjust the shape slightly from any particular drawing independently of $\alpha$.
As a result, the conclusions would be \emph{more optimistic} with respect to detectability than those discussed in Sec.~\ref{sec:results} simply because the GW amplitude scales with the frequency (Eq.~\ref{eq:dimamp}).
The change is offset however by the lower chirp masses, and thus the overall adjustment is modest.
We estimate $\sim \mathcal{O}(50)$ more sources would be visible in this case from the tail, bringing the total detectability fraction to $\sim 0.1\%$.}

{In an effort to depict the limiting values in this work as regards LPT detectability, and to avoid strong astrophysical priors, we opt to consider a single power-law model in the main text even though one may expect a break around $P \sim 80$~min.
Between $\sim 0.05\%$ and $\sim 6\%$ of LPTs may therefore be detectable based on our analysis, with the true number lying somewhere between these two extrema (with the lower limit being closer to $\gtrsim 0.1\%$ if including a GW-dominated transition at $P \sim 80$~min).
}

\section{Fisher matrix analysis} \label{sec:fisher}

Here we present details regarding Fisher matrices and parameter recovery, as used in Sec.~\ref{sec:parameterrecovery}. We closely follow \cite{can12}, \cite{dest20}, and references therein, adopting their notation. We use the software of the Figures of Merit pipeline~\citep{fom_pipeline}. The interferometric response to an impinging GW is determined by a vector $h_{\kappa}$ which is a function of the response functions $F^{+,\times}_{I,II}$ of the antennae and the two independent plus and cross polarisations through \citep{bar04}
\begin{equation}
h_{\kappa}(t)=\frac{\sqrt{3}}{2}\left[F^{+}_{\kappa}(t)h_{+}(t)+F^{\times}_{\kappa}(t)h_{\times}(t)\right],
\end{equation}
where $\kappa=(I,II)$ selects the independent channel of the interferometer, and
\begin{eqnarray}
F^{+}_I &=& \frac{1}{2}(1+\cos^2\theta)\cos(2\phi)\cos(2\psi) \nonumber \\
&&-\cos\theta\sin(2\phi)\sin(2\psi), \\
F^{\times}_I &=& \frac{1}{2}(1+\cos^2\theta)\cos(2\phi)\cos(2\psi)\nonumber \\
&&+\cos\theta\sin(2\phi)\sin(2\psi), \\
F^{+}_{II} &=& \frac{1}{2}(1+\cos^2\theta)\sin(2\phi)\cos(2\psi) \nonumber 
\\&&+\cos\theta\cos(2\phi)\sin(2\psi), \\
F^{\times}_{II} &=& \frac{1}{2}(1+\cos^2\theta)\sin(2\phi)\sin(2\psi) \nonumber \\
&&-\cos\theta\cos(2\phi)\cos(2\psi).
\end{eqnarray}
The angles $\theta$ and $\phi$ here represent spherical polars which characterise the sky location of the source relative to the Solar-system barycentre frame, and $\psi = -2 \pi t /T_{\rm sun}$ where $T_{\rm sun} = 1$~yr is the Earth's orbital period around the Sun.
The components $h_{+,\times}$ relate to the source components \eqref{eq:sourcecomponents}.

Suppose that the data stream witnessed by LISA, $s_{\kappa}(t)$, contains both an ``LPT signal'' $h_{\kappa}(t)$ and Gaussian, stationary noise $n_{\kappa}(t)$. 
We may then write
\begin{equation}
s_\kappa(t) = h_\kappa(t) + n_\kappa(t).
\end{equation}
We also impose the condition that the two data streams are \emph{uncorrelated} and that the PSD of the LISA noise, $S_{\kappa,n}(f)$, is identical in both channels: $S_{I,n}(f) = S_{II,n}(f)$.
This means we can effectively drop the subscript $\kappa$.
The Fourier components of $\boldsymbol{n}$ therefore read
\begin{equation}
\langle \tilde{n}_\alpha (f)\tilde{n}^{\ast}_\beta (f')\rangle = \frac{1}{2} 
\delta^{}_{\alpha\beta} \delta(f-f') S_n(f) ,
\end{equation}
where $\langle \cdot \rangle$ means we take an ``ensemble average'' over the possible realisations. 
Fourier transformed variables are denoted with tildes, and complex conjugates with an asterisk.
For the PSD itself, $S_{n}$, we use the fits from references discussed in the main text (Sec.~\ref{sec:parameterrecovery}).
Thanks to the assumption of Gaussian noise imposed above, the probability that a given signal is present within the data stream is simply given by
\begin{equation} \label{eq:likelihood2}
p(s|h) \sim \exp \left[- \frac {1} {2} (s-h|s-h)\right],
\end{equation}
where $(a |b)$ denotes the inner product on the vector space of signals from expression \eqref{eq:ineerprod}.

The ``best-fitting'' waveform $h$ is typically defined as the one which maximises, in some appropriate sense, the value of $(s|h)$. 
In practice though, a family of waveform templates are used to account for uncertainties.
These depend on a set of parameters, say $\lambda^{i}$; one then searches for those $\lambda^{i}$ such that the probability of a certain noise realisation is maximised \citep{can12}.
Via the central-limit theorem, if the SNR is sufficiently large (chosen as 7 in this paper), the best-fitting $\lambda^{i}_{0}$ should themselves follow a Gaussian distribution that is centered around the ``true'' values. 
Conceptually, if we were to write $\lambda^{i} = \lambda^{i}_{0} + \delta \lambda^{i}$, the task amounts to expanding expression \eqref{eq:likelihood2} around the best-fit values and estimating the impact of the ``perturbations'' $\delta \lambda^{i}$. 
From the likelihood \eqref{eq:likelihood2}, one finds
\begin{equation}
p(\delta \lambda) \sim \exp\left(-\frac{1}{2} \Gamma_{jk} 
\delta\lambda^j\delta\lambda^k\right),
\end{equation}
where
\begin{equation} \label{eq:fisher}
\Gamma_{jk} =\left( \frac{\partial {h}}{\partial \lambda^j} \Big| \frac{\partial {h}}{\partial \lambda^k} \right).
\end{equation}
These $\Gamma$ are precisely the components of the \cite{fish35} information matrix, the inverse of which is the covariance matrix for waveform parameters, viz.
\begin{equation} \label{eq:covariance}
\langle \delta\lambda^j\delta\lambda^k \rangle = \left( \Gamma^{-1}\right)^{jk} \left[1+ \mathcal{O}(\text{SNR}^{-1})\right].
\end{equation}

Finally, using expression \eqref{eq:covariance}, we can estimate the precision $\sigma_{\lambda^{i}}$ with which one may measure the parameter vector $\lambda^{i}$:
\begin{equation} \label{eq:fisherest}
\sigma_{\lambda^{i}} \approx \sqrt{ \left( \Gamma^{-1}\right)^{ii} }. 
\end{equation}
In this context, it is the values \eqref{eq:fisherest} that are presented in the main text (e.g., in Fig.~\ref{fig:fisher}).

\label{lastpage}
\end{document}